\documentclass[floats,floatfix,twocolumn,superscriptaddress,nofootinbib,showpacs]{revtex4-1}

\pdfoutput=1
\usepackage{amsfonts,amssymb,amsmath}
\usepackage[usenames,dvipsnames]{xcolor}
\usepackage[unicode,colorlinks=true,linkcolor=blue,urlcolor=blue,citecolor=gray]{hyperref}
\usepackage[all]{hypcap}
\usepackage{graphicx}
\usepackage{xspace}
\usepackage{tikz}
\usetikzlibrary{shapes,arrows}
\usetikzlibrary{matrix}
\usetikzlibrary{shapes.multipart}
\usetikzlibrary{er,positioning}
\usepackage[normalem]{ulem} 
\usepackage{url}
\usepackage{color,soul}
\usepackage[caption=false]{subfig}
\usepackage[T1]{fontenc}
\usepackage[utf8]{inputenc}
\usepackage{microtype}

\definecolor{forestgreen}{rgb}{0.1,0.5,0.1}

\newcommand{\Caltech}{\affiliation{Theoretical Astrophysics,
    Walter Burke Institute for Theoretical Physics,\\
    California Institute of Technology, Pasadena, CA 91125, USA}}
\newcommand{\Cornell}{\affiliation{Center for Radiophysics and Space
    Research, Cornell University, Ithaca, New York 14853, USA}}
\newcommand{\OleMiss}{\affiliation{Department of Physics and Astronomy, The University of Mississippi, University, MS 38677, USA}}
\newcommand{\Flatiron}{\affiliation{Center for Computational Astrophysics, Flatiron Institute, 162 5th Ave, New York, NY 10010}}

\newcommand{\eps}{\ensuremath{\varepsilon}}

\newcommand{\pd}{\partial}

\newcommand{\nn}{\nonumber}

\newcommand{\txt}[1]{{\textrm{\tiny{#1}}}}
\newcommand{\mpl}{\ensuremath{m_\txt{pl}}}

\newcommand{\pont}{{}^{*}\!RR}

\newcommand{\dual}{\,{}^*\!}

\begin{document}

\title{Numerical relativity simulation of GW150914 beyond general relativity}

\author{Maria Okounkova}
\email{mokounkova@flatironinstitute.org}
\Flatiron \Caltech
\author{Leo C. Stein}
\OleMiss
\author{Jordan Moxon}
\Caltech
\author{Mark A. Scheel}
\Caltech

\author{Saul A. Teukolsky}
\Caltech \Cornell

\date{\today}


\begin{abstract}
We produce the first astrophysically-relevant numerical binary black hole gravitational waveform in a higher-curvature theory of gravity beyond general relativity. We simulate a system with parameters consistent with GW150914, the first LIGO detection, in order-reduced dynamical Chern-Simons gravity, a theory with motivations in string theory and loop quantum gravity. We present results for the leading-order corrections to the merger and ringdown waveforms, as well as the ringdown quasi-normal mode spectrum. We estimate that such corrections may be discriminated in
  detections with signal to noise ratio $\gtrsim 180-240$, with the precise value depending on the dimension of the GR waveform family used in data analysis.
\end{abstract}

\maketitle

\section{Introduction}
\label{sec:Introduction}

Binary black hole mergers, as recently observed by LIGO and Virgo~\cite{LIGOScientific:2018mvr}, probe gravity in its most dynamical, non-linear regime. At some scale, Einstein's theory of general relativity (GR) must break down, and binary black hole (BBH) mergers, by probing strong-field gravity, could potentially contain signatures of beyond-GR effects. A major scientific effort of gravitational wave (GW) astronomy is thus testing general relativity with gravitational wave observations from binary black hole systems~\cite{TheLIGOScientific:2016src, LIGOScientific:2019fpa}.

However, current tests of general relativity are limited to null-hypothesis (assuming GR) and parametrized tests~\cite{Yunes:2016jcc, TheLIGOScientific:2016src}. One future goal is to perform \textit{model-dependent} tests, in which beyond-GR theories of gravity are evaluated with similarly precise methods as those used for GR predictions. A major challenge in this program, however, is the absence of numerical relativity gravitational waveforms in beyond-GR theories through full inspiral, merger, and ringdown. As Yunes et al. argued in~\cite{Yunes:2016jcc}, constraining `physics beyond General Relativity is severely limited by the lack of understanding of the coalescence regime in almost all relevant modified gravity theories'. 

Our goal in this study is to produce the first astrophysically-relevant numerical relativity
binary black hole gravitational waveform in a higher-curvature theory
of gravity.
Specifically, we will focus on dynamical Chern-Simons (dCS) gravity, a beyond-GR effective field theory that adds a scalar field coupled to spacetime curvature to the Einstein-Hilbert action, and has origins in string theory and loop quantum gravity~\cite{Alexander:2009tp, Green:1984sg, Taveras:2008yf, Mercuri:2009zt}. To ensure well-posedness of the evolution equations, we work in an order-reduction scheme, perturbing the dCS scalar field and spacetime metric around general relativity. 

We extend our recent computation of leading order dCS gravitational waveforms for binary black-hole head-on collisions~\cite{MashaHeadOn} to inspiraling systems. Namely, we focus on a simulation with parameters consistent with GW150914, the first LIGO detection, and the loudest so far~\cite{Abbott:2016blz, LIGOScientific:2018mvr}. 

\subsection{Roadmap and conventions} 
\label{sec:conventions}

We give an overview of our methods in Sec.~\ref{sec:methods}, 
and refer the reader to previous papers,~\cite{Okounkova:2017yby, MashaEvPaper, MashaIDPaper, MashaHeadOn}, for technical details. We present the results, including dCS merger and ringdown waveforms for a system consistent with GW150914, in Sec.~\ref{sec:results}. We conclude and discuss future work, including implications for LIGO data analysis, in Sec.~\ref{sec:conclusion}.

We set $G = c = 1$ throughout. Quantities are given in terms of units of $M$, the sum of the Christodoulou masses of the background black holes at a given reference time~\cite{Boyle:2009vi}. Latin letters in the beginning of the alphabet $\{a, b,  c, d \ldots \}$ denote 4-dimensional spacetime indices, and $g_{ab}$ refers to the spacetime metric.

\section{Methods}
\label{sec:methods} 

\subsection{Order-reduction scheme}
\label{sec:order_reduction}

\begin{figure}
  \includegraphics[width=\columnwidth]{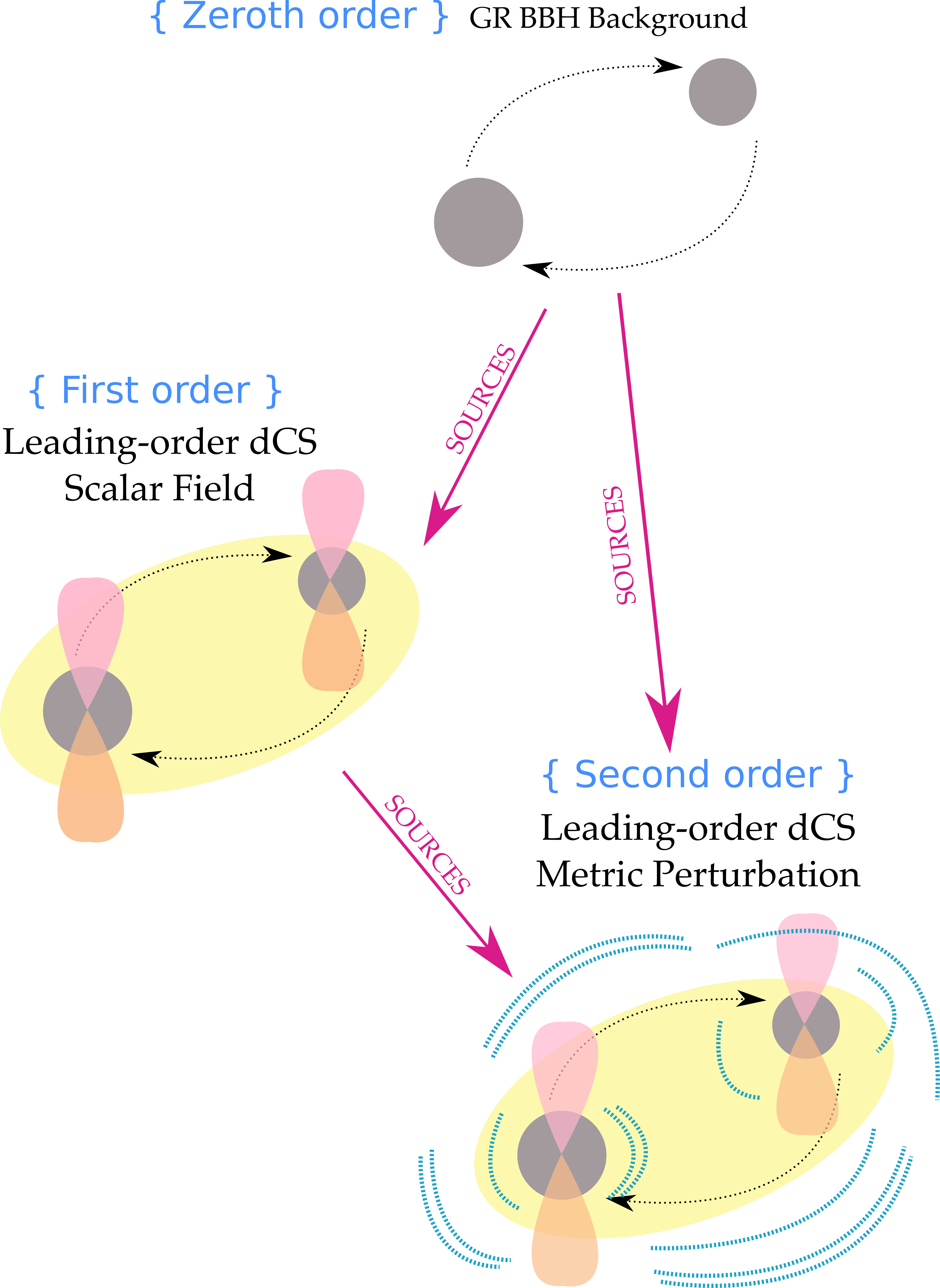}
  \caption{Schematic of the order-reduction scheme, as outlined in
    Sec.~\ref{sec:order_reduction}. We perturb the spacetime metric
    and dCS scalar field around GR in order to ensure well-posedness
    of the evolution equations. At zeroth order, we recover GR, and
    simply have a vacuum GR BBH system. The curvature of this
    background in turn sources the leading-order dCS scalar field
    (coming in at first order, shown in pink and
    yellow) [cf. Eq.~\eqref{eq:CodeScalarField}]. This scalar field and
    the curvature of the GR background then source the leading-order
    dCS correction to the metric (coming in at second order, shown in
    blue) [cf. Eq.~\eqref{eq:CodeMetric}]. It is precisely this
    correction to the spacetime metric that gives the leading-order
    dCS correction to the gravitational waveform.
    }
  \label{fig:order_reduction}
\end{figure}

The action of dynamical Chern-Simons gravity is
\begin{align}
\label{eq:dCSAction}
S \equiv \int d^4 x \sqrt{-g} \left( \frac{\mpl^2}{2} R - \frac{1}{2} (\pd \vartheta)^2 - \frac{\mpl}{8} \ell^2 \vartheta \dual RR \right) \,,
\end{align}
where the first term is the Einstein-Hilbert action of GR, with the Planck
mass denoted by $\mpl$, the second term is a kinetic
term for the (axionic) scalar field $\vartheta$, and the third term
couples $\vartheta$ to spacetime curvature through the parity-odd
Pontryagin density,
\begin{align}
\dual RR \equiv \dual R^{abcd} R_{abcd}\,.
\end{align}
Here, $\dual R^{abcd} = \frac{1}{2} \epsilon^{abef} R_{ef}{}^{cd}$ is
the dual of the Riemann tensor, and $\epsilon_{abcd} \equiv \sqrt{-g}
[abcd]$ is the fully antisymmetric Levi-Civita tensor. The quantity $\ell$ in the third term of Eq.~\eqref{eq:dCSAction} is the dCS coupling constant with dimensions of length, and physically represents the
length scale below which dCS corrections become relevant.

Varying the action Eq.~\eqref{eq:dCSAction} will lead to a set of
field equations which we refer to as the ``full'' equations of dCS.
It is unknown whether full dCS has a well-posed initial value problem
(IVP), though this possibility seems unlikely~\cite{Delsate:2014hba}.
This seems an apparent disqualification for a theory.  However,
we do not take Eq.~\eqref{eq:dCSAction} as an action for an exact
theory.  Rather we assume there is some well-posed underlying UV
theory, and that a low-energy limit gives dCS as an effective field
theory (EFT).

Truncating a high-energy theory to a low-energy EFT has the potential
to introduce extra time derivatives and spurious or runaway
solutions. These spurious solutions are not the low-energy limit of
solutions to the original high-energy theory, and must be eliminated.
Indeed the only consistent way to handle solutions to an EFT are in a
power series in some small parameter $\eps$.  This leads to a
perturbative treatment which reduces the order of the differential
equations, and is considered to be the correct way to excise spurious
solutions (see~\cite{Parker:1993dk, Flanagan:1996gw, Burgess:2014lwa,
  Solomon:2017nlh} for more discussion and examples).  We thus perturb
all fields around GR, as earlier suggested in~\cite{Yunes:2009hc,
  Yagi:2011xp, Stein:2014xba, Okounkova:2017yby}.  This ensures that the
principal symbol of the PDE at each order is the principal symbol of the
generalized harmonic formulation of GR, ensuring a well-posed initial
value problem.

We give details regarding the derivation of the equations of motion for the order-reduction scheme in~\cite{Okounkova:2017yby}. Here, we briefly summarize the important points of the order-reduction construction. We perturb the spacetime metric and dCS scalar field around GR as
\begin{align}
\label{eq:MetricExpansion}
g_{ab} &= g_{ab}^\mathrm{(0)} + \sum_{k = 1}^\infty \eps^k g_{ab}^{(k)} \,, \\
\vartheta &= \sum_{k = 0}^\infty \eps^k \vartheta^{(k)} \,.
\end{align}
Here, terms with superscript ${(0)}$ refer to zeroth-order GR fields,
and $\eps$ is a dimensionless formal parameter counting powers of
$\ell^2$.
We schematically illustrate this order-reduction scheme in Fig.~\ref{fig:order_reduction}. 

We can consistently set $\vartheta^{(0)}=0$ in the background. The leading-order dCS scalar field comes in at first order as $\vartheta^{(1)}$, with the equation of motion
\begin{align}
\label{eq:scalarfield}
    \square^{(0)} \vartheta^{(1)} = \frac{\mpl}{8} \ell^2 \dual RR^{(0)}\,.
\end{align}
Similarly, we can consistently set $g_{ab}^{(1)}=0$.
The leading-order dCS correction to the spacetime metric comes in at second order as $g^{(2)}_{ab}$, with the equation of motion
\begin{align}
\label{eq:SecondOrder}
    \mpl^2 G_{ab}^{(0)} [g_{ab}^{(2)}] = -\mpl \ell^2 C_{ab}^{(1)}[g_{ab}^{(0)}, \vartheta^{(1)}]  + T_{ab}^{(1)}[g_{ab}^{(0)}, \vartheta^{(1)}] \,,
\end{align}
Here, $G_{ab}^{(0)}$ is the linearized
Einstein field equation operator of the background, $T_{ab}^{(1)}$ is
the canonical Klein-Gordon stress energy tensor computed from
$\vartheta^{(1)}$ and $g_{ab}^{(0)}$ (cf. Eq.~(11)
in~\cite{MashaHeadOn}), and $C_{ab}^{(1)}$ is a quantity computed from
the background spacetime curvature and $\vartheta^{(1)}$ (cf. Eq.~(12)
in~\cite{MashaHeadOn}).

We can scale out the $\ell$ dependence (cf.~\cite{Okounkova:2017yby, MashaHeadOn}) by defining new variables
\begin{align}
\label{eq:CodeVariables}
g_{ab}^{(2)} \equiv \frac{(\ell/GM)^4}{8} \Delta g_{ab}\,, \; \; \; \vartheta^{(1)} \equiv \frac{\mpl}{8} (\ell/GM)^2 \Delta \vartheta\,.
\end{align}
Thus, Eqs.~\eqref{eq:scalarfield} and~\eqref{eq:SecondOrder} become
\begin{align}
\label{eq:CodeScalarField}
   \square^{(0)} \Delta \vartheta &= \dual RR^{(0)}\,, \\
   \label{eq:CodeMetric}
G_{ab}^{(0)} [\Delta g_{ab}] &= -C_{ab}^{(1)}[\Delta \vartheta] + \frac{1}{8}T_{ab}^{(1)}[\Delta \vartheta]\,.
\end{align}
Along with the nonlinear equations for the background metric,
Eqs.~\eqref{eq:CodeScalarField} and~\eqref{eq:CodeMetric} are numerically co-evolved to obtain the leading-order dCS correction to the gravitational waveforms. 

\subsection{Inspiral secular growth}
\label{sec:secular_growth}

In the order-reduction scheme, the motion of the black holes is governed by the GR background. The GR background sources a dCS scalar field which in turn sources a dCS metric deformation (cf. Fig.~\ref{fig:order_reduction}), but this perturbation to the spacetime does \textit{not} back-react onto the GR background. Thus the trajectories of the black holes, and hence the rate of inspiral, are purely determined by GR.

In full dCS gravity, however, the black holes will inspiral faster than in GR because they lose energy to scalar radiation and because the gravitational-radiation energy loss is modified from GR. There is thus a discrepancy between the rate
of BBH inspiral in the order-reduction scheme and in the ``full'' dCS
theory, which leads to secular breakdown of perturbation
theory~\cite{MR538168}. In particular, this effect occurs on the
radiation-reaction timescale, which governs the motion of the black
holes towards one another.

The inspiral portion of the leading-order dCS modification to the gravitational waveform will thus be contaminated by secular effects. Removing these effects, in particular through renormalization, is the subject of future work that we discuss in Appendix~\ref{sec:secular_appendix}. Another potential avenue for extending the accuracy of a perturbative scheme to the full inspiral includes stitching to known post-Newtonian expressions for the  dCS modification to the waveform in the early inspiral~\cite{Yagi:2011xp}.

In this paper, we will \textit{focus on the merger and ringdown} portions of the leading-order dCS correction to the gravitational waveform. To mitigate secular effects from inspiral, our goal is to start the evolution of the dCS fields $\Delta \vartheta$ and $\Delta g_{ab}$ as close to merger as possible. However, starting a binary black hole simulation close to merger creates a host of problems involving initial data and initial transients commonly called junk radiation~\cite{Lovelace:2008hd, Pfeiffer:2007yz, Buonanno:2010yk}. We thus evolve a standard BBH GR background simulation, and \textit{ramp on} the source terms for $\Delta \vartheta$ and $\Delta g_{ab}$ (cf. Eqs.~\eqref{eq:CodeScalarField} and~\eqref{eq:CodeMetric}) starting at some later time $t_s$. We give more details about the ramp functions in Appendix~\ref{sec:ramp_appendix}.

\subsection{Technical details}
We use the same evolution framework as our previous head-on collisions work~\cite{MashaHeadOn}, to which we refer the reader for technical details. All of the computations are performed using the Spectral Einstein Code (SpEC), which uses pseudo-spectral methods and thus guarantees exponential convergence in all of the fields. The main technical change between the treatment of head-on collisions and inspiraling mergers is tracking the orbiting black holes on the computational grid, which is performed using the methods of~\cite{Hemberger:2012jz}. 


\section{GW150914 Results}
\label{sec:results}

In this section, we present the results of performing a binary black hole simulation using the methods of Sec.~\ref{sec:methods} for a system consistent with GW150914. In particular, we compute the leading-order dCS modification to the merger and ringdown waveforms. All presented waveforms are asymptotic, computed from an expansion in $1/R$ for extraction radius $R$~\cite{Boyle:2009vi}, and hence should have near-field effects removed. 

\subsection{Simulation parameters}
\label{sec:simulation}

While there is a distribution of mass and spin parameters consistent
with GW150914~\cite{TheLIGOScientific:2016wfe, Kumar:2018hml}, we
choose to use the parameters of SXS:BBH:0305, as given in the
Simulating eXtreme Spacetimes (SXS) catalog~\cite{SXSCatalog}. This
simulation was used in Fig. 1 of the GW150914 detection
paper~\cite{Abbott:2016blz}, as well a host of follow-up
studies~\cite{Lovelace:2016uwp, Bhagwat:2017tkm, Giesler:2019uxc}. The
configuration has initial dimensionless spins $\chi_A = 0.330 \hat{z}$
and $\chi_B = -0.440 \hat{z}$, aligned and anti-aligned with the
orbital angular momentum. The dominant GR spherical harmonic modes of
the gravitational radiation for this system are $(l,m)=(2,\pm2)$. The
system has initial masses of $0.5497 \,M$ and $0.4502 \,M$, leading to
a mass ratio of $1.221$. The initial eccentricity is
$\sim 8 \times 10^{-4}$. The black holes merge at $t = 2533.8 \,M$,
forming a common horizon after $\sim 23$ orbits. The remnant has final
Christodoulou mass $0.9525 \,M$ and dimensionless spin $0.692$ purely
in the $\hat{z}$ direction.

Note that the mass ratio and spins $(q, \chi_A, \chi_B)$ that best fit GW150914 in GR may be different from the best-fit dCS-modified waveform parameters, $(q', \chi_A', \chi_B', \ell/GM)$. In this initial paper, we focus on just one set of background parameters, but future work includes performing studies in this extended parameter space to explore posterior reconstructions and degeneracies, among other topics, as we outline in Sec.~\ref{sec:conclusion}.

\subsection{dCS merger waveform}
\label{sec:merger}

\begin{figure}
  \includegraphics[width=\columnwidth]{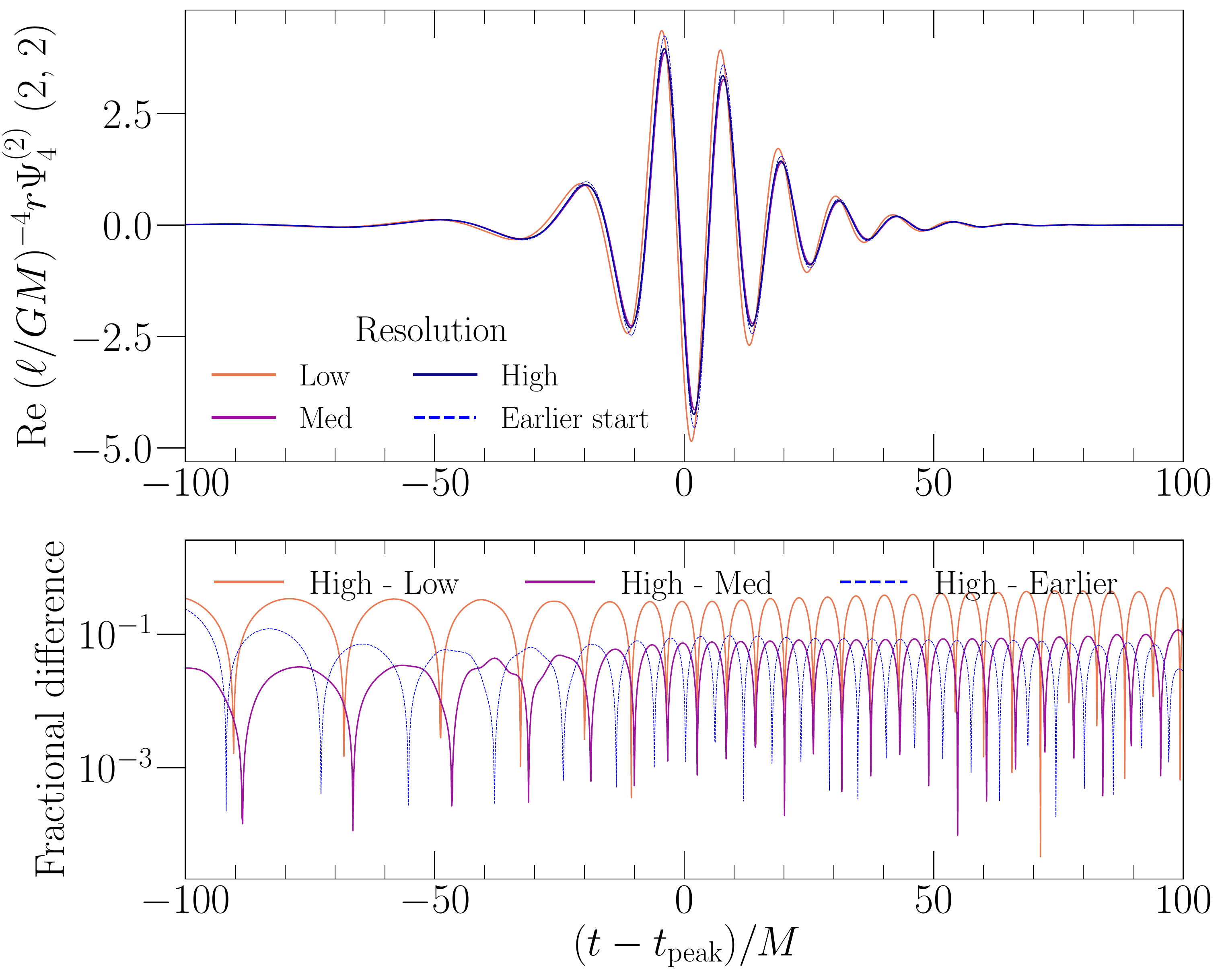}
  \caption{Leading-order dCS correction to the gravitational waveform
    for a system with parameters consistent with GW150914
    (cf. Sec.~\ref{sec:simulation}). As we discuss in Sec.~\ref{sec:observed_secular}, this
    merger gravitational waveform is not contaminated by secular effects. In the upper panel, we show the waveform correction for low,
    medium, and high numerical resolutions. The curve labeled `Earlier start' shows a waveform with a slightly earlier dCS start time ($25\,M$ before, cf. Sec.~\ref{sec:observed_secular}), which lies between the different numerical resolution waveforms. In the lower panel, we show the fractional differences (normalized by the complex amplitudes) between the low, medium, and high resolutions, showing that the waveform converges with numerical resolution. We show the fractional difference between the highest resolution waveform and the `Earlier start' waveform, and find that this lies within the numerical error bounds
    of the merger waveform. We thus conclude that there is no significant amplitude
    difference caused by an earlier start time. Hence, our merger waveform
    is not contaminated by secular effects.
    Note that all of the
    waveforms are presented with the dCS coupling $(\ell/GM)^4$ scaled
    out, and this factor must be reintroduced for the results to be
    physically meaningful.
    }
  \label{fig:hPsi4}
\end{figure}

The main result of this paper is Fig.~\ref{fig:hPsi4}, which shows the leading-order dCS correction to the gravitational waveform for GW150914 during merger. We focus on the dominant $(l,m)=(2,\pm 2)$ modes of the gravitational radiation (though we have access to gravitational wave modes from $l = 2$ to $l = 8$), as these are the most important modes for GW150914 observation and analysis~\cite{Abbott:2016blz, TheLIGOScientific:2016src, TheLIGOScientific:2016wfe}. The $(2,-2)$ mode is consistent with being the complex conjugate
  of the $(2,2)$ mode so we only present results for the latter.

In order to make this waveform useful for LIGO data analysis, we must reintroduce the dCS coupling parameter $\ell$, and present the full, second-order accurate dCS curvature waveform as (cf. Eq.~\eqref{eq:CodeVariables}),
\begin{align}
\label{eq:total_psi4}
    \Psi_4 = \Psi_4^{(0)} + \Psi_4^{(2)} \,,\\
    \Psi_4^{(2)} \equiv \frac{(\ell/GM)^4}{8} \Delta \Psi_4 \,,
\end{align}
where $\Delta \Psi_4$ is the linearized curvature
perturbation (with the dCS coupling scaled out) computed from the 
variable $\Delta g_{ab}$ (cf. Eq.~\eqref{eq:CodeMetric}).  Thus, given one numerical simulation, we
can generate a family of waveforms parametrized by $\ell$ by simply
multiplying and adding using Eq.~\eqref{eq:total_psi4}.

\begin{figure}
  \includegraphics[width=\columnwidth]{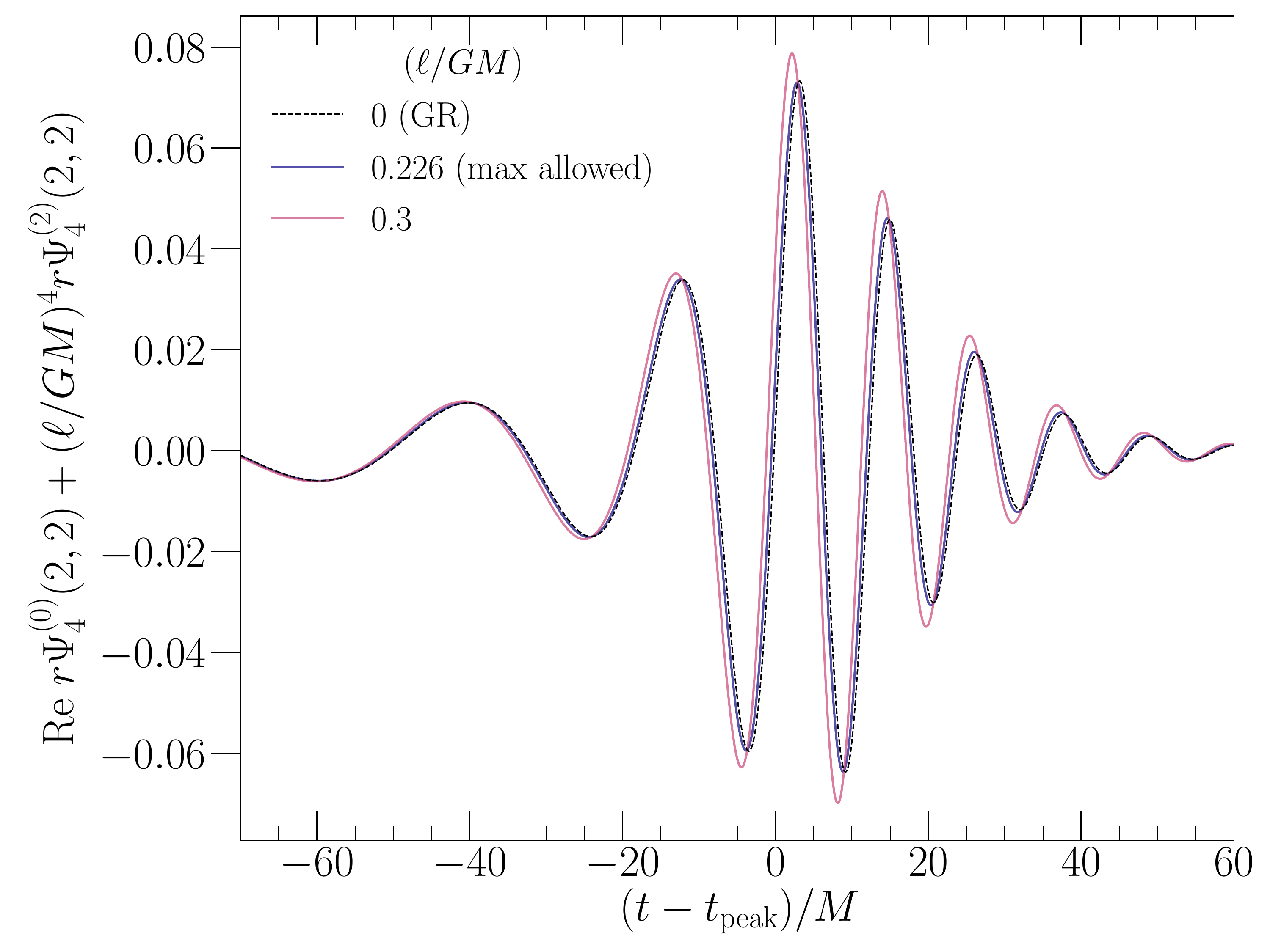}
  \caption{Second-order accurate dCS gravitational waveforms, for three choices of dCS coupling constant, $\ell/GM$. We add the
    leading-order dCS correction to the gravitational waveform (from
    Fig.~\ref{fig:hPsi4}) to the background GR gravitational waveform
    of the system, to give 
    the total dCS waveform [cf.~\eqref{eq:total_psi4}].
    The value
    $\ell/GM = 0$ corresponds to GR, with no dCS modifications. The
    value $\ell/GM = 0.226$ corresponds to the largest-allowed value
    for the perturbative scheme to be valid
    (cf. Sec.~\ref{sec:instantaneous_validity}). The $\ell/GM = 0.3$
    curve is included to visually emphasize the shape of alteration
    provided by the dCS correction.
  }
  \label{fig:Together}
\end{figure}

\begin{figure}
  \includegraphics[width=\columnwidth]{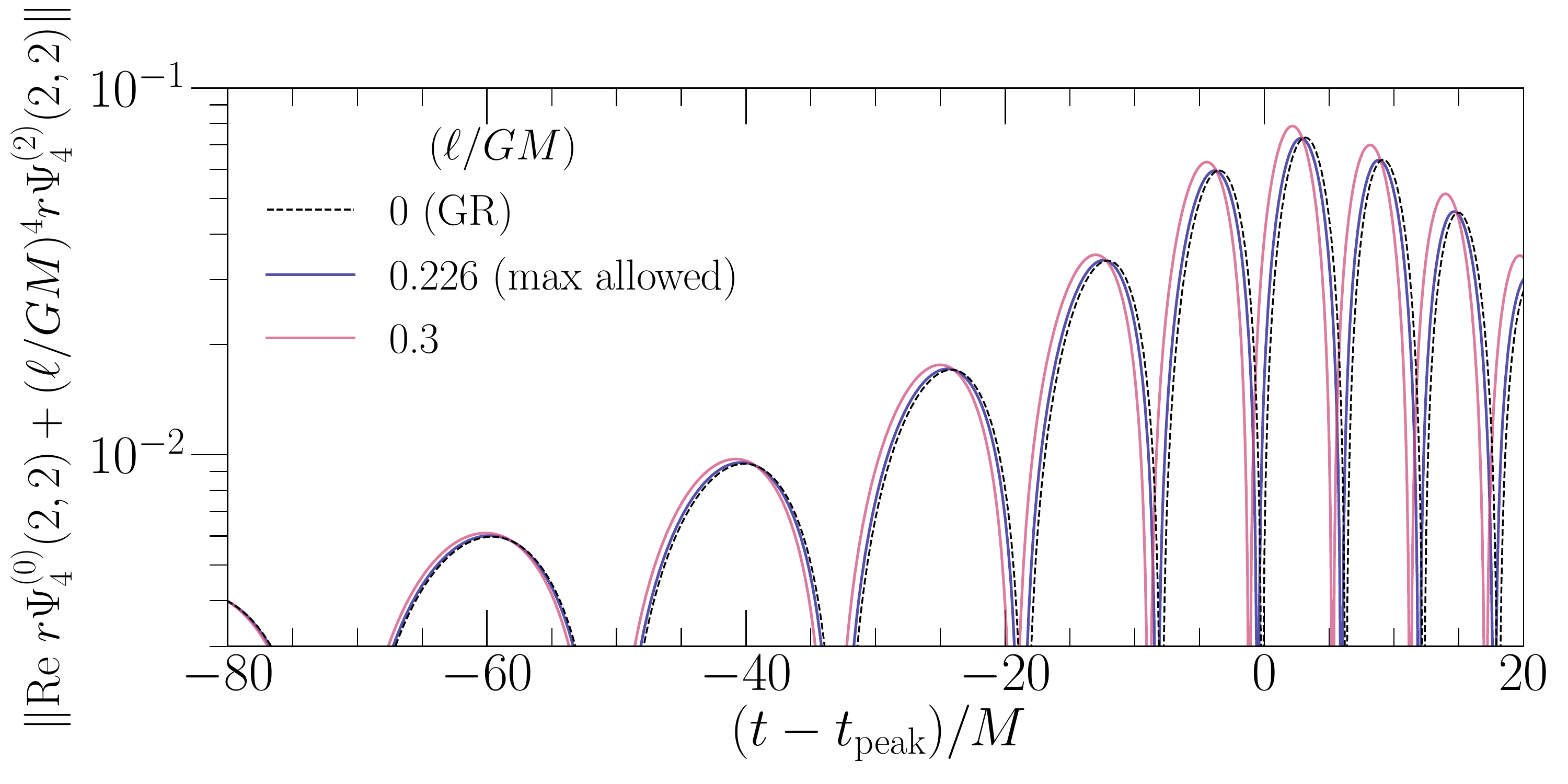}
  \caption{Same as Fig.~\ref{fig:Together}, but on a logarithmic scale to better show the phase difference between the GR and dCS corrected waveforms.}
  \label{fig:Together_log}
\end{figure}

In Fig.~\ref{fig:Together}, we show the total, second-order accurate
dCS gravitational waveform for three choices of coupling parameter
$\ell/GM$.  When comparing two different waveforms with each other, we are allowed to make an overall time shift,
but the background $\Psi_{4}^{(0)}$ and correction
  $\Psi_{4}^{(2)}$ of the same waveform may not be shifted relative to each other.  We
have shifted the time axis on $\Psi_4^{(0)}$ and $\Psi_4^{(2)}$ to the
peak time of the real part of $\Psi_4^{(0)}$.  We see a
\emph{time-dependent} modification in the phase of the total
  waveform in Fig.~\ref{fig:Together}.  We plot the results on a logarithmic scale in
Fig.~\ref{fig:Together_log} for clarity.  The time dependence of the phase shift is crucial for the dCS correction to be non-degenerate
  with the background GR waveform.  The sign of the shift is consistent with the intuition
that a dCS-corrected binary inspirals more quickly, since energy
  can be lost through the scalar field.  Thus the waveform should have an
earlier merger than the pure GR waveform.

\subsection{Mismatch}
\label{sec:mismatch}

Let us now compute the mismatch between the dCS waveform and the corresponding GR waveform. Note that we do not optimize over different background GR parameters, but rather compute the mismatch for the waveform presented in Sec.\ref{sec:merger} for $\ell/GM = 0$ (GR) and $\ell/GM = 0.226$ (the maximal allowed value by the instantaneous regime of validity in Sec.~\ref{sec:instantaneous_validity}). Using the methods of Sec. VII of~\cite{Varma:2018mmi}, we compute the strain mismatch with the Advanced-LIGO design sensitivity noise curve~\cite{LIGOPSD}. For an optimally-oriented binary with total mass $68 M_\odot$, we compute a mismatch, optimizing over time and phase shift, of $8.6 \times 10^{-5}$. This is an approximate upper bound on the
  mismatch, which we will analyze more carefully in future work.
  
We can estimate a minimum SNR necessary to
  distinguish between GR and the dCS-corrected waveforms.  We follow
  the distinguishability criterion Eq.~(G13)
  of~\cite{Chatziioannou:2017tdw},
  \begin{align}
    \label{eq:distinguish-SNR}
    \mathcal{M} \gtrsim \frac{D}{2\text{ SNR}^{2}}
    \,,
  \end{align}
  where $D$ counts the number of parameters in the model used by data analysis (cf. Appendix G in~\cite{Chatziioannou:2017tdw} for a derivation of this expression).  Using this
  criterion, we find a lower limit on the minimum SNR for
  distinguishability,
  \begin{align}
    \text{SNR} \gtrsim 80 \sqrt{D}
    \,.
  \end{align}

  What should the value of $D$, the number of parameters in the data analysis waveform model, be? The largest value for $D$ for a circular binary would be $D = 9$ for parameters $\vec \lambda = \{m_1, m_2, \vec \chi_1, \vec \chi_2, \ell\}$, where $m_\mathrm{1,2}$ are the masses of the two holes (note that this can also be reparametrized in terms of the chirp mass $\mathcal{M}$ and mass ratio $q$), and $\vec \chi_\mathrm{1,2}$ are the 3-dimensional spins of the black holes. The dCS coupling parameter $\ell$ is an additional parameter in a beyond-GR analysis. Gravitational wave data analysis often does not make use of the full 8-dimensional GR parameter space, however. In the LIGO parameter estimation study for GW150914~\cite{TheLIGOScientific:2016wfe}, two waveform models were used: EOBNR~\cite{Taracchini:2013rva} and precessing IMRPhenom, (cf.~\cite{Khan:2015jqa}). The EOBNR model has 4 intrinsic GR parameters $\vec \lambda = \{m_1, m_2, \chi_\mathrm{eff}, \chi_p\}$, where $\chi_\mathrm{eff}$ corresponds to the component of the spins along the orbital angular momentum, and $\chi_p$ corresponds to the perpendicular component. This would correspond to $D = 5$ for our purposes. The IMRPhenom model, meanwhile, uses 4 of the 6 spin degrees of freedom, which corresponding to $D = 7$. See also the parameter estimation in GWTC-1~\cite{LIGOScientific:2018mvr}, which provides an updated analysis. Perhaps the smallest value of $D$ would be $D = 3$, corresponding to $\vec \lambda = \{\mathcal{M}, \chi_\mathrm{eff}, \ell\}$, as $\mathcal{M}$ and $\chi_\mathrm{eff}$ are parameters LIGO can most robustly measure (cf.~\cite{LIGOScientific:2018mvr}). Thus, let us quote values for $D = \{3, 5, 7, 9\}$, giving 
  \begin{align}
    \text{SNR} \{D = 3, 5, 7, 9\} \gtrsim \{138, 179, 211, 240\}\,.
  \end{align}
Let us quote the detectability SNR as the range $\text{SNR}  \gtrsim 180-240$, with the precise value depending on the dimension of the parameter estimation model. It is beyond the scope of this study to consider which class of models should be used specifically, as well as the biases introduced by only including a subspace of the full 9 dimensional parameter space (see~\cite{Purrer:2019jcp} for recent work in this direction).

\subsection{dCS ringdown waveforms}
\label{sec:qnm}

\begin{figure}
  \includegraphics[width=\columnwidth]{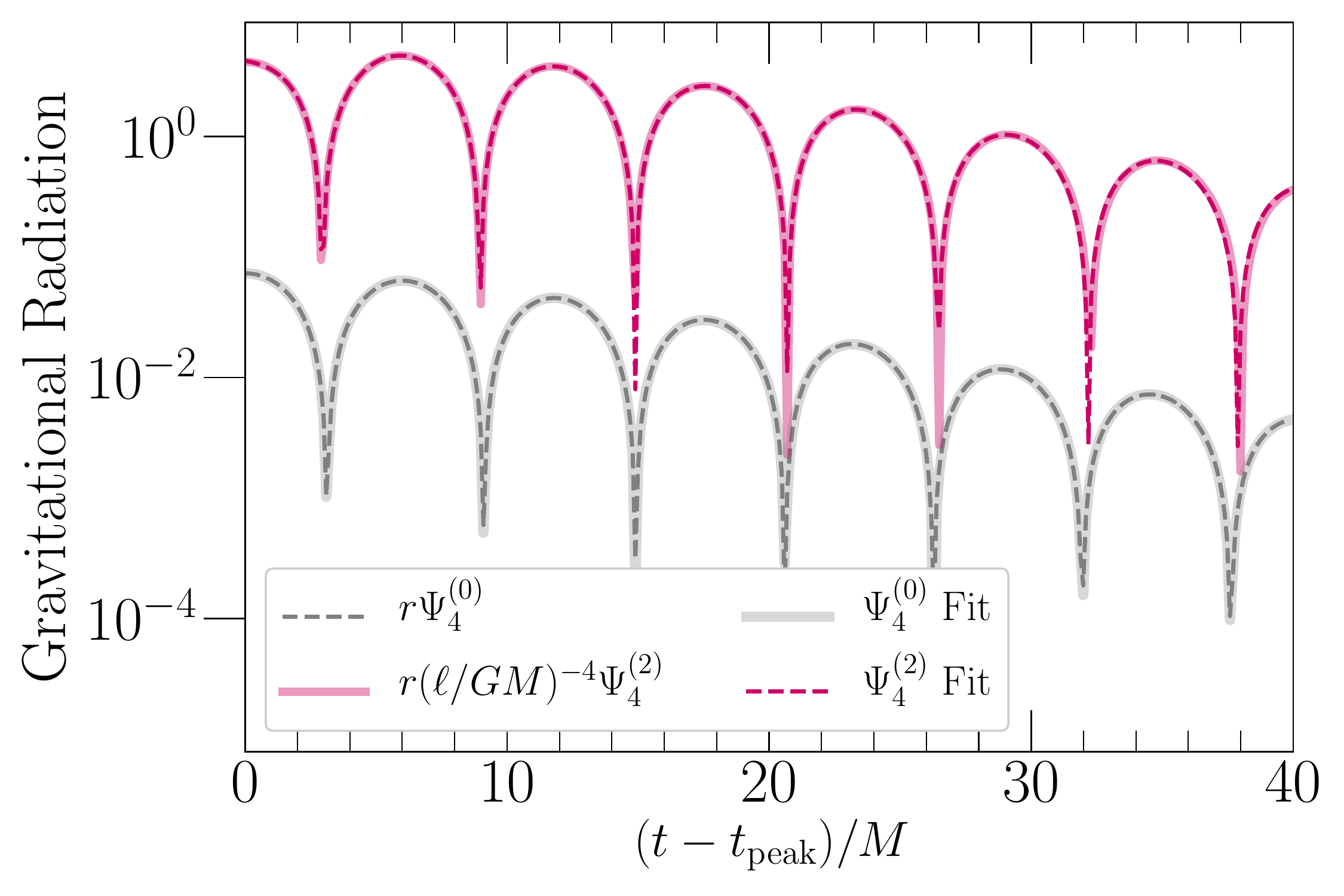}
  \caption{Quasinormal mode fits for  the post-merger spectrum of $r\Psi_4^{(2)}$ (dashed pink curve), the leading-order dCS gravitational radiation. We show the dominant $(2,2)$ mode of the radiation, fit to the three least-damped overtones. The solid colored curve corresponds to the real part of $r\Psi_4^{(2)}$. For reference, we have plotted the real part of $r\Psi_4^{(0)}$ in dashed gray. This is similar to Fig.~7 of~\cite{MashaHeadOn}, which was done for a head-on collision in order-reduced dCS. }
  \label{fig:RDFit}
\end{figure}

Let us now repeat the dCS ringdown analysis of~\cite{MashaHeadOn} to compute the leading-order modifications to the GW150914 quasi-normal mode (QNM) spectrum. We fit the dominant $(2,2)$ mode. Using three overtones, we are able to fit all the way to the peak of $\Psi_4$, in line with the results of~\cite{Giesler:2019uxc, Isi:2019aib}. We show the ringdown modification fit in Fig.~\ref{fig:RDFit} (cf. Fig.~7 of~\cite{MashaHeadOn}). We give the leading-order dCS modifications to the QNM frequency and damping time, with the dCS coupling parameter scaled out in Table~\ref{tab:qnms}.

\begin{table}[htb!]
\begin{center}
  \begin{tabular}{ c | c | c }
    
     $n$ & $(\ell/GM)^{-4}\omega^{(2)}_{(2,2,n)} M_\mathrm{f} $  & $(\ell/GM)^{-4} \tau^{(2)}_{(2,2,n)} / M_\mathrm{f}$ \\ \hline

0 & $-0.437 \pm 0.03$ & $-8.13 \pm 0.25$  \\ \hline
1 & $3.92 \pm 0.14$ & $220.1 \pm 6.5$\\ \hline 
2 & $-1.54 \pm 0.04$ & $146.9 \pm 6.4$ \\ \hline

  \end{tabular}
\caption{Fitted QNM parameters for the post-merger gravitational radiation for the GW150914 simulation considered in this study (cf. Sec.~\ref{sec:simulation}). Each row corresponds to one of the three dominant overtones of the $(2,2)$ mode of the radiation. The quantity $\omega^{(2)}_{(2,2,n)} M_\mathrm{f} $ is the leading-order dCS modification of the QNM frequency (multiplied by the final mass $M_\mathrm{f}$), while $\tau^{(2)}_{(2,2,n)}/ M_\mathrm{f}$ is the leading-order dCS modification to the damping time (divided by the final mass $M_\mathrm{f}$). In each case, the dCS coupling parameter $(\ell/GM)^{4}$ is scaled out. For this simulation, the final mass is $M_\mathrm{f} = 0.9525\,M$, and the final dimensionless spin is $\chi_f = 0.692$, purely in the $\hat{z}$ direction.}
\label{tab:qnms}
\end{center}
\end{table}

\subsection{dCS secular growth during inspiral}
\label{sec:observed_secular}

\begin{figure}
  \includegraphics[width=\columnwidth]{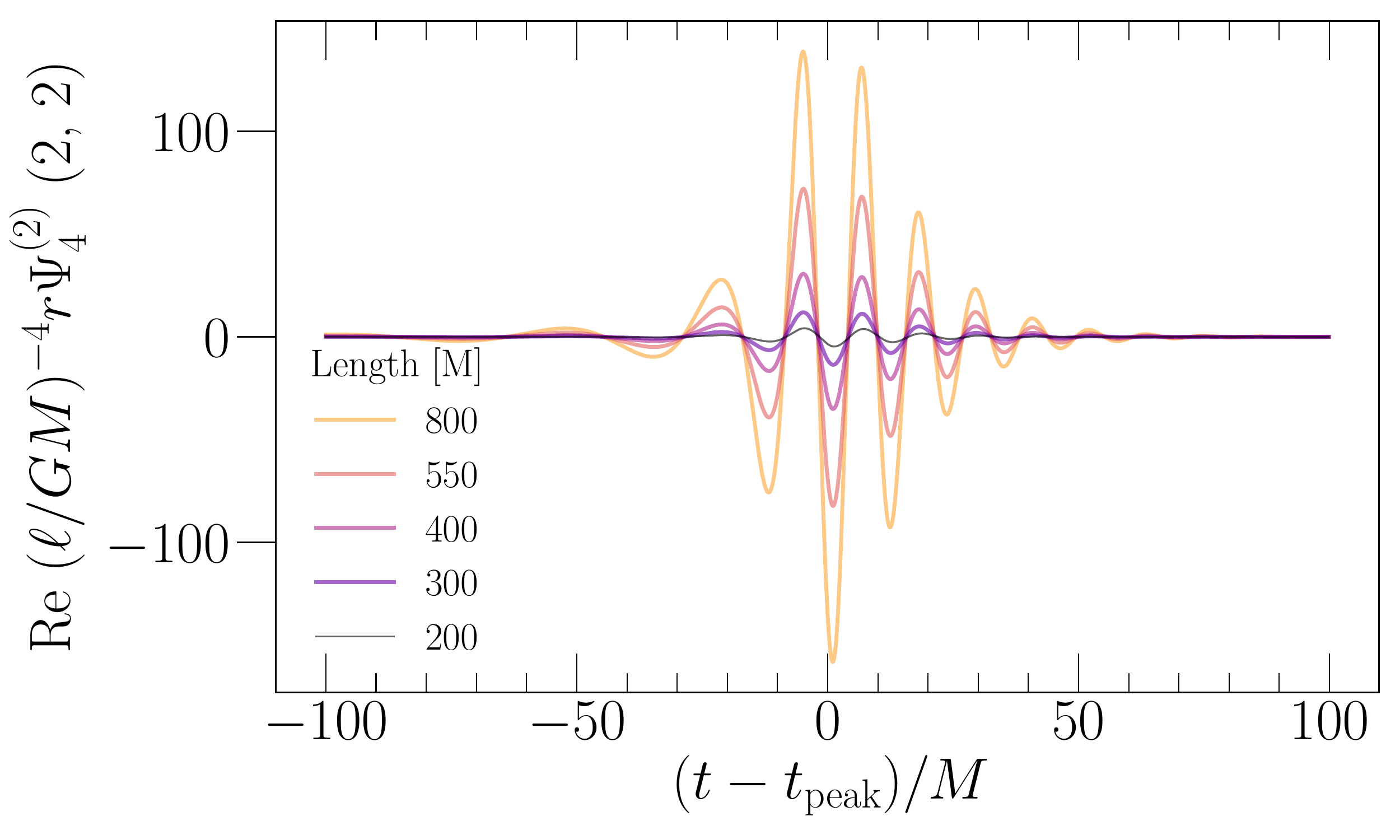}
  \caption{Secular growth in leading-order dCS gravitational waveforms as function of the length of the waveform. Each colored curve corresponds to a simulation with a different start time for the dCS fields (as discussed in Sec.~\ref{sec:secular_growth}), with the same GR background simulation for each. We label each curve by the time difference between the peak of the waveform and the start time of ramping on the dCS field (minus the ramp time). We see that simulations with earlier dCS start times have higher amplitudes at merger, having had more time to accumulate secular growth.}
  \label{fig:SecularhPsi4}
\end{figure}

As predicted in Sec.~\ref{sec:secular_growth}, we see secular growth during the inspiral in the leading-order dCS waveforms, $\Psi_4^{(2)}$. In Fig.~\ref{fig:SecularhPsi4}, we show the results for  $\Psi_4^{(2)}$ for simulations with various dCS start times (and hence different times over which to accumulate secular growth). We see that the longest dCS simulations have the largest amplitudes of  $\Psi_4^{(2)}$ at merger, and the shortest dCS simulations have comparably small amplitudes. Thus, when physically interpreting the inspiral results, we must remove the secular growth accumulated over the inspiral phase, which is work in progress that we discuss in Appendix~\ref{sec:secular_appendix}.

\begin{figure}
  \includegraphics[width=\columnwidth]{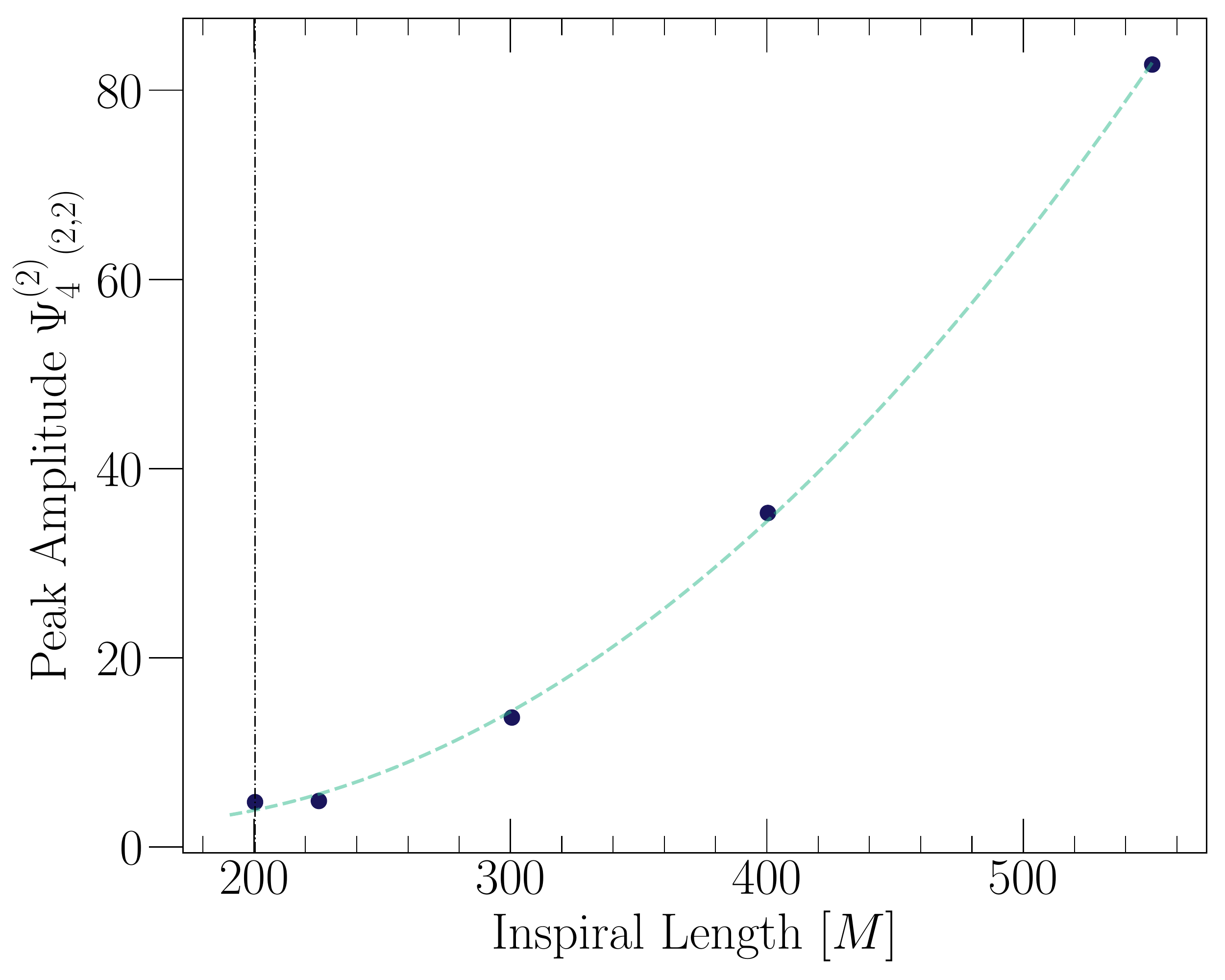}
  \caption{Peak amplitude of the dCS correction to the gravitational waveform as a function of inspiral length. We show the length relative to the peak of the waveform (as in Fig.~\ref{fig:SecularhPsi4}). The dashed black vertical line corresponds to the length of the dCS merger simulation we present in this paper. The peak amplitude serves as a measure of the amount of secular growth in the waveform (cf. Fig.~\ref{fig:SecularhPsi4}). We see that the peak amplitude increases quadratically with inspiral length (as shown by the quadratic fit in dashed green). 
  }
  \label{fig:PeakAmplitude}
\end{figure}

To trust our waveform, we must ensure it is not contaminated by
  secular effects. We performed 5 additional simulations with different start times
  for turning on the dCS correction.
In Fig.~\ref{fig:PeakAmplitude}, we look at the peak amplitude of the
waveform as a function of inspiral length.  We see that the secular
growth, as reflected in the peak amplitude, behaves quadratically with
the length of the dCS simulation.
The waveform we presented in Secs.~\ref{sec:merger} and
  \ref{sec:qnm} correspond to an inspiral length of $200 M$.  Since
  this is near the quadratic minimum, it shows minimal sensitivity to
  length and minimal secular contamination.
In Fig.~\ref{fig:hPsi4} we have also plotted waveforms with
  length $225 M$.  Note that
the difference in amplitudes between
the two waveforms is smaller than the difference in amplitude
we see between different numerical resolutions of waveforms with inspiral length
of $200 M$.
Thus, we conclude that the merger waveform presented in
Sec.~\ref{sec:merger} is not contaminated by secular effects.

\subsection{Regime of validity}

\label{sec:instantaneous_validity}

Besides the previously-discussed secular breakdown, there is also an
instantaneous regime of validity.
Since we work in a perturbative scheme, at each instant there is a
finite radius of convergence in $\ell/GM$.
We estimate this value using the formalism in~\cite{Stein:2014xba,
  MashaHeadOn}, by comparing the size of the leading-order correction
$\Delta g_{ab}$ to the background metric $g_{ab}^{(0)}$.
If the correction becomes comparable to the background, it is no longer justified to neglect the
omitted higher-order terms in the expansion. For the waveform
presented in Fig.~\ref{fig:hPsi4}, the maximum allowed value of
$\ell/GM$, as a function of time, attains its minimum value at merger,
where the strong-field effects are greatest, and hence the dCS metric
deformation is largest.
Here we find $(\ell/GM)_{\max} \approx 0.226$.
If we take a total mass $M \sim 68 M_\odot$ (consistent with
GW150914~\cite{TheLIGOScientific:2016wfe}), we can use our perturbative treatment to make
  self-consistent calculations up to $\ell_{\max} \approx 23\text{ km}$.
  That is, if data analysis using this waveform leads to a constraint
  tighter than $\ell < \ell_{\max}$, the use of perturbation
  theory was consistent, and the constraint is valid --- even for
  ``full'' dCS theory.

\section{Conclusions and future work}
\label{sec:conclusion}

In this study, we have produced the first astrophysically-relevant numerical relativity binary
black hole gravitational waveform in a higher-curvature theory of
gravity. 
We have focused on dynamical Chern-Simons gravity, a
quadratic gravity theory with origins in string theory and loop
quantum gravity~\cite{Alexander:2009tp}, extending our previous
results for binary black hole head on collisions in dCS to inspiraling
systems~\cite{MashaHeadOn}.

We have focused on a BBH system with parameters consistent with
GW150914, the first LIGO detection~\cite{TheLIGOScientific:2016wfe,
  Lovelace:2016uwp} (cf. Sec.~\ref{sec:simulation}). In
Sec.~\ref{sec:merger}, we presented the leading-order dCS correction
to the merger gravitational waveform, with minimal secular effects.  In Sec.~\ref{sec:qnm}, we repeated our
quasi-normal mode analysis presented in~\cite{MashaHeadOn}, analyzing
the leading-order dCS correction to the ringdown waveform, and
extracting the corresponding modifications to the frequencies and
damping times of the quasi-normal modes.

In Sec.~\ref{sec:observed_secular}, we showed the presence of secular growth during the inspiral portion of the leading-order dCS gravitational waveform, as theoretically predicted in Sec.~\ref{sec:secular_growth}. While we address possible avenues for removing this secular growth in Appendix~\ref{sec:secular_appendix}, we focused in this study on the merger and ringdown portions of the waveform. 

Our ultimate goal is to make these beyond-GR waveforms useful for LIGO
and Virgo tests of general relativity~\cite{TheLIGOScientific:2016src,
  LIGOScientific:2019fpa}. While a natural conclusion would be to
generate enough beyond-GR waveforms to fill the BBH parameter space,
build a surrogate model, and use this for model-dependent parameter
estimation (cf.~\cite{Kumar:2018hml}), an important first step is to
study the degeneracies between beyond-GR waveforms and waveforms in
pure general relativity. We have found non-degeneracy with GR in the
limit of infinite signal to noise ratio in the dCS black hole
shadow~\cite{MashaIDPaper}, and in the quasi-normal mode
spectrum~\cite{MashaHeadOn}, but a realistic analysis must include
detector noise. We will inject our beyond-GR waveforms
into LIGO noise and compute posteriors recovered using present LIGO parameter estimation and testing-GR methods~\cite{Aasi:2013jjl, Cornish:2014kda,
  TheLIGOScientific:2016wfe, LIGOScientific:2019fpa}. This in turn serves as a \textit{degeneracy study}, testing the degeneracy of a dCS-corrected waveform with GR waveforms with different parameters in the presence of LIGO noise.

Additionally, our methods are fully general~\cite{MashaEvPaper},
and thus can be used for Einstein dilaton Gauss-Bonnet (EDGB) gravity,
another higher curvature theory.
While simulations of the
leading order EDGB scalar field on a BBH background have been
performed~\cite{Witek:2018dmd}, we can go one order higher and obtain
the leading-order EDGB correction to the gravitational waveform. This
is forthcoming work, and we have already demonstrated the
leading-order numerical stability of rotating BHs in EDGB with our
methods~\cite{Okounkova:2019zep}.
EDGB, however, has dipolar radiation during the inspiral, with the
leading-order post-Newtonian correction to the inspiral entering at
$-1$ PN order relative to GR (dCS enters at $2$ PN relative to
GR)~\cite{Yagi:2011xp}. Thus secular effects should be larger in EDGB than in dCS.
  However, if one can control these secular effects, or show minimal
  contamination, EDGB should enjoy more stringent constraints than
  dCS.

\acknowledgements
We thank Vijay Varma for computing the mismatch in Sec.~\ref{sec:mismatch}. We thank Katerina Chatziioannou, Chad Galley, Francois Hebert for helpful discussions. We thank Dante Iozzo for useful comments on this manuscript. This work was supported in part by the Sherman Fairchild Foundation, and NSF grants PHY-1708212 and PHY-1708213 at Caltech and PHY-1606654 at Cornell. The Flatiron Institute is supported by the Simons Foundation.
LCS acknowledges support from award No.~80NSSC19M0053 to the MS NASA EPSCoR RID Program.
All computations were performed on the Wheeler cluster at Caltech, which is supported by the Sherman Fairchild Foundation and by Caltech. All simulations are performed using the Spectral Einstein Code (SpEC)~\cite{SpECwebsite}. 

\appendix

\section{Strategies for removing secular growth}
\label{sec:secular_appendix}
A common feature of perturbative approximations to near-oscillatory
dynamical systems is the tendency to develop unbounded secular divergence from
the exact physical solution~\cite{MR538168}.
Of course, once such growth 
reaches a magnitude such that the secularly 
growing solution competes with the small parameter of the expansion, the
solution should not be trusted at all.
However, prior to that point, the secular growth represents an expanding
deviation from the hypothetical exact solution.
This growth degrades the precision of the approximate solution.
Because of secular growth, the scale of the error term in the expansion of the solution of 
Eqs.~\eqref{eq:CodeScalarField} and~\eqref{eq:CodeMetric} should no longer be estimated as $\mathcal{O}((\ell/GM)^4)$, but instead should be estimated as
$\mathcal{O}(\Delta)$, where
\begin{equation} \label{eq:DeltaError}
\Delta = \left(\frac{\ell}{GM}\right)^4 \frac{t}{T},
\end{equation}
and $T$ is the radiation-reaction timescale.

These secularly growing perturbed solutions are not intrinsically
erroneous, but instead represent a nontrivial evolution of the
background parameters of the system.  For the dCS expansion relevant
to this paper, those background parameters can be thought of as the
initial data parameters of the inspiral~\cite{Galley:2016zee}, such as
the mass and spin of the black holes, or as parameters relevant to the
waveform, such as the amplitude and phase.  Different
parameterizations are similarly valid, but give rise to distinct
details in how the mitigation strategies are formulated.  The secular
growth does present a practical problem for solutions, though, as
numerical results will eventually drift from the perturbative domain
of validity (cf. Sec \ref{sec:instantaneous_validity}), and fail to
approximate the true dynamics at long timescales.

In the dCS perturbed system, the secular drift is most obvious in the
frequency parameter of the binary, which then manifests itself as an 
approximately quadratic in time $\sim t^2/T^2$ drift of the orbital phase parameter
(see Sec.~\ref{sec:secular_growth}).
The body of this paper describes the robust results for near merger and
ringdown of the binary that can be obtained by enforcing that the $\Delta$ 
of Eq.~\eqref{eq:DeltaError} is within an acceptable tolerance level throughout 
the simulation, which is validated in Sec.~\ref{sec:instantaneous_validity}.

To illustrate why such effects arise, consider the generic perturbed
equations of motion approximating a linear second-order hyperbolic
differential equation $D(g) = 0$:
\begin{subequations} \label{eq:GenericGrowthEom}
\begin{align}
D^{(0)}(g^{(0)}) &= 0, \label{eq:GenericGrowthEomA} \\
D^{(0)}(g^{(1)}) &= D^{(1)}(g^{(0)}), \label{eq:GenericGrowthEomB}
\end{align}
\end{subequations}
where $g = g^{(0)} + \epsilon g^{(1)} + \mathcal{O}(\epsilon^2)$ and 
$D = D^{(0)} + \epsilon D^{(1)} + \mathcal{O}(\epsilon^2)$.
Consider a set of homogeneous solutions $g^{(0)}(C_i)$ to
Eq.~\eqref{eq:GenericGrowthEomA} parameterized by constants $C_i$.
Consider then a linearized differential operator $D^{(1)}$ with properties
such that
\begin{equation}
  D^{(1)}(g^{(0)}) = \sum_i \beta^i
  \frac{\delta g^{(0)}(\hat C_i) }{\delta \hat C_i},
\end{equation}
for a set of coefficients $\beta^i$
approximately constant over some sufficiently short time and 
new set of approximate constants $\hat C_i$. Then
Eq.~\eqref{eq:GenericGrowthEomB} is solved by
$g^{(1)} = \alpha(x^\mu) g^{(0)}(t \beta_i + \hat C_i)$, for
$\alpha$ determined by the coefficients of the derivatives in
$D^{(0)}$. The secular growth can then be seen in the linear
dependence of the arguments in the solution of $g^{(1)}$.

The statements of the previous paragraph can be extended to more generic 
systems and parameters, and can be used to determine the sets
of functions in an appropriate decomposition of the space of possible
right-hand sides of Eq.~\eqref{eq:GenericGrowthEomB} that give rise to secular growth.
The exploration of the nature of the growing solutions and methods
to mitigate the secular growth are the topics of several mathematical 
publications, e.g.~\cite{Kevorkian,Kunihiro:1995zt}, and references therein.

The problems of secular growth in approximate solutions to binary
black hole inspirals have been noticed, and at least partially
addressed, in other calculational contexts.  Particularly,
Post-Newtonian approximation~\cite{Will:2016pgm} and self-force black
hole perturbation theory~\cite{Pound:2005fs} encounter these
challenges when applied to long-duration inspirals. A simple solution
to the problem is to `stitch' together results from different starting
points of the evolutions~\cite{Pound:2007th, Warburton:2011fk}.  The
analogous method for the dCS computation in this paper would be to
transition between the inspiral predicted by different start times
$t_S$ of the dCS computation.  Such simple methods, however, will fail
to recover the phase accuracy we seek for gravitational wave
predictions.  A full solution to the secular growth problem is to
infer the slow evolution of the parameters of the background solution
(e.g. phase and amplitude drift of the waveform), and use that slow
evolution to adjust the background solution while performing
appropriate alterations to the perturbed solution to ensure that the
full equations of motion remain satisfied to the desired precision.
The presentation of these types of methods can be found in various
sources~\cite{Kunihiro:1995zt, Galley:2016zee, Hinderer:2008dm,
  Pound:2015wva}, and some of the most promising techniques are
referred to in the literature as `multiscale' methods and `dynamical
renormalization group' methods.

The main challenges in implementing a method to bring the phase evolution
back into the background solution are: 1) inferring the rate of the phase
evolution from a given secularly growing perturbative solution, and 2) applying
that correction while maintaining desired precision.
The task of performing the background correction is made all the more
difficult without the formulaic luxury of exact analytic backgrounds that 
prior work has enjoyed.
We intend to handle in future work the task of extracting the phase drift from
a secularly growing solution by a well-chosen fit to a linear combination of
candidate homogeneous waveform of the $g^{(2)}_{ab}$ perturbed metric and a slowly
drifting numerical solution.

An adiabatic solution should then be available by evolving the slowly
varying parameters of the pure-GR background solution through the space of
perturbative solutions. In forthcoming work, we address the feasibility of
performing that adiabatic evolution as a post-processing step to waveforms
obtained from a family of short-duration approximate dCS evolutions such 
as those presented in this paper.

\section{Ramping on dCS source terms}
\label{sec:ramp_appendix}

As discussed in Sec.~\ref{sec:secular_growth}, our goal is to start the dCS
simulations as close to merger as possible, in order to avoid secular
growth effects from the inspiral. We thus aim to ramp on the source
terms for $\Delta g_{ab}$ and $\Delta \vartheta$ [as governed by
Eqs.~\eqref{eq:CodeScalarField} and~\eqref{eq:CodeMetric}] at some
start time $t_s$ close to merger.

Consider some ramp function of the form 
\begin{align}
  f(t) =
  \begin{cases}
    0 & t < t_{s} \\
    F(t) & t_{s} \leq t \leq t_{s} + t_{\text{ramp}} \\
    1 & t_{s} + t_{\text{ramp}} < t \,,
  \end{cases}
\end{align}
where $F(t)$ smoothly ramps from 0 to 1 with a specified number of
derivatives matching at each endpoint.  The required number of
derivatives depends on the order of the integration scheme.
Here $t_\mathrm{ramp}$ is the
characteristic ramp time of $F(t)$. We ramp on the dCS fields by
replacing $\ell^2 \to f(t) \ell^2$ in the evolution equations, and
ramp on the scalar field source as
\begin{align}
\label{eq:RampedField}
    \square^{(0)} \Delta \theta &= f(t) \pont^{(0)}\,.
\end{align}
We similarly ramp on the dCS metric deformation as 
\begin{align}
\label{eq:RampedMetric}
    G_{ab}^{(0)}[\Delta g_{ab}] &= - f(t) C_{ab}^{(1)} [\Delta \theta] + \frac{1}{8} T_{ab}^{(1)}[\Delta \theta] \,.
\end{align}
For the above equation, recall that in Eq.~\eqref{eq:SecondOrder}, there is a factor of $\ell^2$ in the $C_{ab}$ term, and thus for the scheme to be consistent we must include a factor of $f(t)$ in front of $C_{ab}^{(1)}$. 

\begin{figure}[tb]
  \includegraphics[width=\columnwidth]{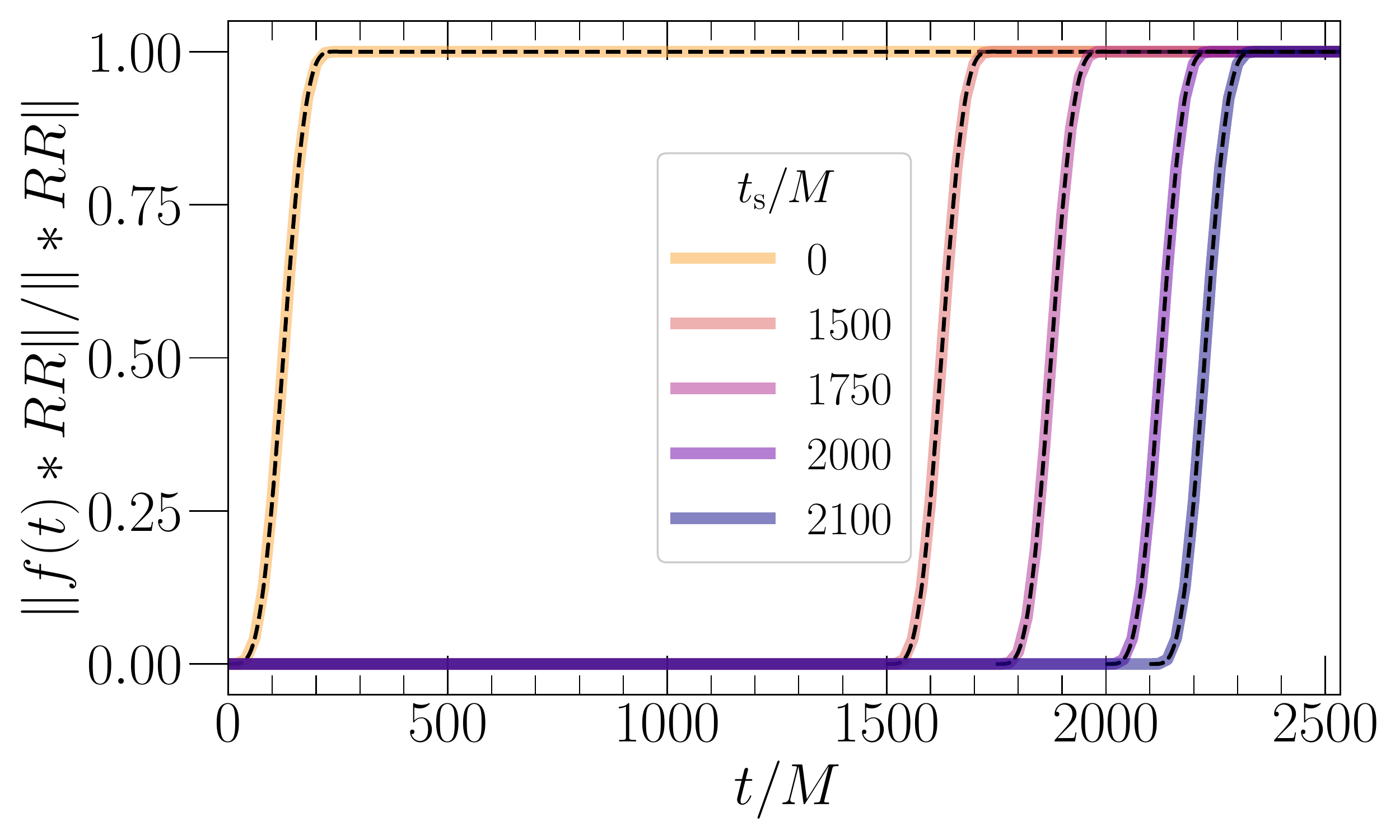}
  \caption{Normalized ramped scalar field source term as a function of coordinate time of the simulation. Each curve represents a different dCS simulation start time $t_\mathrm{s}$. The source is ramped on for $t_\mathrm{ramp} = 250\,M$ until it achieves the value of $\dual  RR$ on the spacetime. We plot the L2 norm of the source term $\dual  RR f(t)$ divided by the value of $\dual  RR$ on the spacetime, and see this goes to 1 once the ramping on is complete. Over each colored curve, we plot a dashed black curve with the analytical functional form of our ramp function, Eq.~\eqref{eq:ramp}, to check consistency.
  }
  \label{fig:RampedSource}
\end{figure}

In practice, we choose a function of the form (between $t_s$ and $t_s + t_\mathrm{ramp}$)
\begin{align}
\label{eq:ramp}
    t_*  &\equiv (t - t_\mathrm{s})/t_\mathrm{ramp} \,, \\
    \nn F(t) &= t_*^5  (126 + t_* (-420 + t_* (540 + t_* (-315 + 70 t_*))))\,.
\end{align} 
We plot the ramped scalar field source term $f(t) \dual RR$ in Fig.~\ref{fig:RampedSource} to show the character of this function.

\bibliography{dCS_paper}

\begin{thebibliography}{54}%
\makeatletter
\providecommand \@ifxundefined [1]{%
 \@ifx{#1\undefined}
}%
\providecommand \@ifnum [1]{%
 \ifnum #1\expandafter \@firstoftwo
 \else \expandafter \@secondoftwo
 \fi
}%
\providecommand \@ifx [1]{%
 \ifx #1\expandafter \@firstoftwo
 \else \expandafter \@secondoftwo
 \fi
}%
\providecommand \natexlab [1]{#1}%
\providecommand \enquote  [1]{``#1''}%
\providecommand \bibnamefont  [1]{#1}%
\providecommand \bibfnamefont [1]{#1}%
\providecommand \citenamefont [1]{#1}%
\providecommand \href@noop [0]{\@secondoftwo}%
\providecommand \href [0]{\begingroup \@sanitize@url \@href}%
\providecommand \@href[1]{\@@startlink{#1}\@@href}%
\providecommand \@@href[1]{\endgroup#1\@@endlink}%
\providecommand \@sanitize@url [0]{\catcode `\\12\catcode `\$12\catcode
  `\&12\catcode `\#12\catcode `\^12\catcode `\_12\catcode `\%12\relax}%
\providecommand \@@startlink[1]{}%
\providecommand \@@endlink[0]{}%
\providecommand \url  [0]{\begingroup\@sanitize@url \@url }%
\providecommand \@url [1]{\endgroup\@href {#1}{\urlprefix }}%
\providecommand \urlprefix  [0]{URL }%
\providecommand \Eprint [0]{\href }%
\providecommand \doibase [0]{http://dx.doi.org/}%
\providecommand \selectlanguage [0]{\@gobble}%
\providecommand \bibinfo  [0]{\@secondoftwo}%
\providecommand \bibfield  [0]{\@secondoftwo}%
\providecommand \translation [1]{[#1]}%
\providecommand \BibitemOpen [0]{}%
\providecommand \bibitemStop [0]{}%
\providecommand \bibitemNoStop [0]{.\EOS\space}%
\providecommand \EOS [0]{\spacefactor3000\relax}%
\providecommand \BibitemShut  [1]{\csname bibitem#1\endcsname}%
\let\auto@bib@innerbib\@empty
\bibitem [{\citenamefont {Abbott}\ \emph
  {et~al.}(2019{\natexlab{a}})\citenamefont {Abbott} \emph
  {et~al.}}]{LIGOScientific:2018mvr}%
  \BibitemOpen
  \bibfield  {author} {\bibinfo {author} {\bibfnamefont {B.~P.}\ \bibnamefont
  {Abbott}} \emph {et~al.} (\bibinfo {collaboration} {LIGO Scientific,
  Virgo}),\ }\href {\doibase 10.1103/PhysRevX.9.031040} {\bibfield  {journal}
  {\bibinfo  {journal} {Phys. Rev.}\ }\textbf {\bibinfo {volume} {X9}},\
  \bibinfo {pages} {031040} (\bibinfo {year} {2019}{\natexlab{a}})},\ \Eprint
  {http://arxiv.org/abs/1811.12907} {arXiv:1811.12907 [astro-ph.HE]}
  \BibitemShut {NoStop}%
\bibitem [{\citenamefont {Abbott}\ \emph
  {et~al.}(2016{\natexlab{a}})\citenamefont {Abbott} \emph
  {et~al.}}]{TheLIGOScientific:2016src}%
  \BibitemOpen
  \bibfield  {author} {\bibinfo {author} {\bibfnamefont {B.~P.}\ \bibnamefont
  {Abbott}} \emph {et~al.} (\bibinfo {collaboration} {Virgo, LIGO
  Scientific}),\ }\href {\doibase 10.1103/PhysRevLett.116.221101,
  10.1103/PhysRevLett.121.129902} {\bibfield  {journal} {\bibinfo  {journal}
  {Phys. Rev. Lett.}\ }\textbf {\bibinfo {volume} {116}},\ \bibinfo {pages}
  {221101} (\bibinfo {year} {2016}{\natexlab{a}})},\ \bibinfo {note} {[Erratum:
  Phys. Rev. Lett.121,no.12,129902(2018)]},\ \Eprint
  {http://arxiv.org/abs/1602.03841} {arXiv:1602.03841 [gr-qc]} \BibitemShut
  {NoStop}%
\bibitem [{\citenamefont {Abbott}\ \emph
  {et~al.}(2019{\natexlab{b}})\citenamefont {Abbott} \emph
  {et~al.}}]{LIGOScientific:2019fpa}%
  \BibitemOpen
  \bibfield  {author} {\bibinfo {author} {\bibfnamefont {B.~P.}\ \bibnamefont
  {Abbott}} \emph {et~al.} (\bibinfo {collaboration} {LIGO Scientific,
  Virgo}),\ }\href@noop {} {\  (\bibinfo {year} {2019}{\natexlab{b}})},\
  \Eprint {http://arxiv.org/abs/1903.04467} {arXiv:1903.04467 [gr-qc]}
  \BibitemShut {NoStop}%
\bibitem [{\citenamefont {Yunes}\ \emph {et~al.}(2016)\citenamefont {Yunes},
  \citenamefont {Yagi},\ and\ \citenamefont {Pretorius}}]{Yunes:2016jcc}%
  \BibitemOpen
  \bibfield  {author} {\bibinfo {author} {\bibfnamefont {N.}~\bibnamefont
  {Yunes}}, \bibinfo {author} {\bibfnamefont {K.}~\bibnamefont {Yagi}}, \ and\
  \bibinfo {author} {\bibfnamefont {F.}~\bibnamefont {Pretorius}},\ }\href
  {\doibase 10.1103/PhysRevD.94.084002} {\bibfield  {journal} {\bibinfo
  {journal} {Phys. Rev.}\ }\textbf {\bibinfo {volume} {D94}},\ \bibinfo {pages}
  {084002} (\bibinfo {year} {2016})},\ \Eprint
  {http://arxiv.org/abs/1603.08955} {arXiv:1603.08955 [gr-qc]} \BibitemShut
  {NoStop}%
\bibitem [{\citenamefont {Alexander}\ and\ \citenamefont
  {Yunes}(2009)}]{Alexander:2009tp}%
  \BibitemOpen
  \bibfield  {author} {\bibinfo {author} {\bibfnamefont {S.}~\bibnamefont
  {Alexander}}\ and\ \bibinfo {author} {\bibfnamefont {N.}~\bibnamefont
  {Yunes}},\ }\href {\doibase 10.1016/j.physrep.2009.07.002} {\bibfield
  {journal} {\bibinfo  {journal} {Phys. Rept.}\ }\textbf {\bibinfo {volume}
  {480}},\ \bibinfo {pages} {1} (\bibinfo {year} {2009})},\ \Eprint
  {http://arxiv.org/abs/0907.2562} {arXiv:0907.2562 [hep-th]} \BibitemShut
  {NoStop}%
\bibitem [{\citenamefont {Green}\ and\ \citenamefont
  {Schwarz}(1984)}]{Green:1984sg}%
  \BibitemOpen
  \bibfield  {author} {\bibinfo {author} {\bibfnamefont {M.~B.}\ \bibnamefont
  {Green}}\ and\ \bibinfo {author} {\bibfnamefont {J.~H.}\ \bibnamefont
  {Schwarz}},\ }\href {\doibase 10.1016/0370-2693(84)91565-X} {\bibfield
  {journal} {\bibinfo  {journal} {Phys. Lett.}\ }\textbf {\bibinfo {volume}
  {149B}},\ \bibinfo {pages} {117} (\bibinfo {year} {1984})}\BibitemShut
  {NoStop}%
\bibitem [{\citenamefont {Taveras}\ and\ \citenamefont
  {Yunes}(2008)}]{Taveras:2008yf}%
  \BibitemOpen
  \bibfield  {author} {\bibinfo {author} {\bibfnamefont {V.}~\bibnamefont
  {Taveras}}\ and\ \bibinfo {author} {\bibfnamefont {N.}~\bibnamefont
  {Yunes}},\ }\href {\doibase 10.1103/PhysRevD.78.064070} {\bibfield  {journal}
  {\bibinfo  {journal} {Phys. Rev.}\ }\textbf {\bibinfo {volume} {D78}},\
  \bibinfo {pages} {064070} (\bibinfo {year} {2008})},\ \Eprint
  {http://arxiv.org/abs/0807.2652} {arXiv:0807.2652 [gr-qc]} \BibitemShut
  {NoStop}%
\bibitem [{\citenamefont {Mercuri}\ and\ \citenamefont
  {Taveras}(2009)}]{Mercuri:2009zt}%
  \BibitemOpen
  \bibfield  {author} {\bibinfo {author} {\bibfnamefont {S.}~\bibnamefont
  {Mercuri}}\ and\ \bibinfo {author} {\bibfnamefont {V.}~\bibnamefont
  {Taveras}},\ }\href {\doibase 10.1103/PhysRevD.80.104007} {\bibfield
  {journal} {\bibinfo  {journal} {Phys. Rev.}\ }\textbf {\bibinfo {volume}
  {D80}},\ \bibinfo {pages} {104007} (\bibinfo {year} {2009})},\ \Eprint
  {http://arxiv.org/abs/0903.4407} {arXiv:0903.4407 [gr-qc]} \BibitemShut
  {NoStop}%
\bibitem [{\citenamefont {Okounkova}\ \emph
  {et~al.}(2019{\natexlab{a}})\citenamefont {Okounkova}, \citenamefont {Stein},
  \citenamefont {Scheel},\ and\ \citenamefont {Teukolsky}}]{MashaHeadOn}%
  \BibitemOpen
  \bibfield  {author} {\bibinfo {author} {\bibfnamefont {M.}~\bibnamefont
  {Okounkova}}, \bibinfo {author} {\bibfnamefont {L.~C.}\ \bibnamefont
  {Stein}}, \bibinfo {author} {\bibfnamefont {M.~A.}\ \bibnamefont {Scheel}}, \
  and\ \bibinfo {author} {\bibfnamefont {S.~A.}\ \bibnamefont {Teukolsky}},\
  }\href {\doibase 10.1103/PhysRevD.100.104026} {\bibfield  {journal} {\bibinfo
   {journal} {Phys. Rev. D}\ }\textbf {\bibinfo {volume} {100}},\ \bibinfo
  {pages} {104026} (\bibinfo {year} {2019}{\natexlab{a}})}\BibitemShut
  {NoStop}%
\bibitem [{\citenamefont {Abbott}\ \emph
  {et~al.}(2016{\natexlab{b}})\citenamefont {Abbott} \emph
  {et~al.}}]{Abbott:2016blz}%
  \BibitemOpen
  \bibfield  {author} {\bibinfo {author} {\bibfnamefont {B.~P.}\ \bibnamefont
  {Abbott}} \emph {et~al.} (\bibinfo {collaboration} {Virgo, LIGO
  Scientific}),\ }\href {\doibase 10.1103/PhysRevLett.116.061102} {\bibfield
  {journal} {\bibinfo  {journal} {Phys. Rev. Lett.}\ }\textbf {\bibinfo
  {volume} {116}},\ \bibinfo {pages} {061102} (\bibinfo {year}
  {2016}{\natexlab{b}})},\ \Eprint {http://arxiv.org/abs/1602.03837}
  {arXiv:1602.03837 [gr-qc]} \BibitemShut {NoStop}%
\bibitem [{\citenamefont {Okounkova}\ \emph {et~al.}(2017)\citenamefont
  {Okounkova}, \citenamefont {Stein}, \citenamefont {Scheel},\ and\
  \citenamefont {Hemberger}}]{Okounkova:2017yby}%
  \BibitemOpen
  \bibfield  {author} {\bibinfo {author} {\bibfnamefont {M.}~\bibnamefont
  {Okounkova}}, \bibinfo {author} {\bibfnamefont {L.~C.}\ \bibnamefont
  {Stein}}, \bibinfo {author} {\bibfnamefont {M.~A.}\ \bibnamefont {Scheel}}, \
  and\ \bibinfo {author} {\bibfnamefont {D.~A.}\ \bibnamefont {Hemberger}},\
  }\href {\doibase 10.1103/PhysRevD.96.044020} {\bibfield  {journal} {\bibinfo
  {journal} {Phys. Rev.}\ }\textbf {\bibinfo {volume} {D96}},\ \bibinfo {pages}
  {044020} (\bibinfo {year} {2017})},\ \Eprint
  {http://arxiv.org/abs/1705.07924} {arXiv:1705.07924 [gr-qc]} \BibitemShut
  {NoStop}%
\bibitem [{\citenamefont {Okounkova}\ \emph
  {et~al.}(2019{\natexlab{b}})\citenamefont {Okounkova}, \citenamefont
  {Scheel},\ and\ \citenamefont {Teukolsky}}]{MashaEvPaper}%
  \BibitemOpen
  \bibfield  {author} {\bibinfo {author} {\bibfnamefont {M.}~\bibnamefont
  {Okounkova}}, \bibinfo {author} {\bibfnamefont {M.~A.}\ \bibnamefont
  {Scheel}}, \ and\ \bibinfo {author} {\bibfnamefont {S.~A.}\ \bibnamefont
  {Teukolsky}},\ }\href {\doibase 10.1103/PhysRevD.99.044019} {\bibfield
  {journal} {\bibinfo  {journal} {Phys. Rev.}\ }\textbf {\bibinfo {volume}
  {D99}},\ \bibinfo {pages} {044019} (\bibinfo {year} {2019}{\natexlab{b}})},\
  \Eprint {http://arxiv.org/abs/1811.10713} {arXiv:1811.10713 [gr-qc]}
  \BibitemShut {NoStop}%
\bibitem [{\citenamefont {Okounkova}\ \emph
  {et~al.}(2019{\natexlab{c}})\citenamefont {Okounkova}, \citenamefont
  {Scheel},\ and\ \citenamefont {Teukolsky}}]{MashaIDPaper}%
  \BibitemOpen
  \bibfield  {author} {\bibinfo {author} {\bibfnamefont {M.}~\bibnamefont
  {Okounkova}}, \bibinfo {author} {\bibfnamefont {M.~A.}\ \bibnamefont
  {Scheel}}, \ and\ \bibinfo {author} {\bibfnamefont {S.~A.}\ \bibnamefont
  {Teukolsky}},\ }\href {\doibase 10.1088/1361-6382/aafcdf} {\bibfield
  {journal} {\bibinfo  {journal} {Class. Quant. Grav.}\ }\textbf {\bibinfo
  {volume} {36}},\ \bibinfo {pages} {054001} (\bibinfo {year}
  {2019}{\natexlab{c}})},\ \Eprint {http://arxiv.org/abs/1810.05306}
  {arXiv:1810.05306 [gr-qc]} \BibitemShut {NoStop}%
\bibitem [{\citenamefont {Boyle}\ and\ \citenamefont
  {Mroue}(2009)}]{Boyle:2009vi}%
  \BibitemOpen
  \bibfield  {author} {\bibinfo {author} {\bibfnamefont {M.}~\bibnamefont
  {Boyle}}\ and\ \bibinfo {author} {\bibfnamefont {A.~H.}\ \bibnamefont
  {Mroue}},\ }\href {\doibase 10.1103/PhysRevD.80.124045} {\bibfield  {journal}
  {\bibinfo  {journal} {Phys. Rev.}\ }\textbf {\bibinfo {volume} {D80}},\
  \bibinfo {pages} {124045} (\bibinfo {year} {2009})},\ \Eprint
  {http://arxiv.org/abs/0905.3177} {arXiv:0905.3177 [gr-qc]} \BibitemShut
  {NoStop}%
\bibitem [{\citenamefont {Delsate}\ \emph {et~al.}(2015)\citenamefont
  {Delsate}, \citenamefont {Hilditch},\ and\ \citenamefont
  {Witek}}]{Delsate:2014hba}%
  \BibitemOpen
  \bibfield  {author} {\bibinfo {author} {\bibfnamefont {T.}~\bibnamefont
  {Delsate}}, \bibinfo {author} {\bibfnamefont {D.}~\bibnamefont {Hilditch}}, \
  and\ \bibinfo {author} {\bibfnamefont {H.}~\bibnamefont {Witek}},\ }\href
  {\doibase 10.1103/PhysRevD.91.024027} {\bibfield  {journal} {\bibinfo
  {journal} {Phys. Rev.}\ }\textbf {\bibinfo {volume} {D91}},\ \bibinfo {pages}
  {024027} (\bibinfo {year} {2015})},\ \Eprint {http://arxiv.org/abs/1407.6727}
  {arXiv:1407.6727 [gr-qc]} \BibitemShut {NoStop}%
\bibitem [{\citenamefont {Parker}\ and\ \citenamefont
  {Simon}(1993)}]{Parker:1993dk}%
  \BibitemOpen
  \bibfield  {author} {\bibinfo {author} {\bibfnamefont {L.}~\bibnamefont
  {Parker}}\ and\ \bibinfo {author} {\bibfnamefont {J.~Z.}\ \bibnamefont
  {Simon}},\ }\href {\doibase 10.1103/PhysRevD.47.1339} {\bibfield  {journal}
  {\bibinfo  {journal} {Phys. Rev.}\ }\textbf {\bibinfo {volume} {D47}},\
  \bibinfo {pages} {1339} (\bibinfo {year} {1993})},\ \Eprint
  {http://arxiv.org/abs/gr-qc/9211002} {arXiv:gr-qc/9211002 [gr-qc]}
  \BibitemShut {NoStop}%
\bibitem [{\citenamefont {Flanagan}\ and\ \citenamefont
  {Wald}(1996)}]{Flanagan:1996gw}%
  \BibitemOpen
  \bibfield  {author} {\bibinfo {author} {\bibfnamefont {E.~E.}\ \bibnamefont
  {Flanagan}}\ and\ \bibinfo {author} {\bibfnamefont {R.~M.}\ \bibnamefont
  {Wald}},\ }\href {\doibase 10.1103/PhysRevD.54.6233} {\bibfield  {journal}
  {\bibinfo  {journal} {Phys. Rev.}\ }\textbf {\bibinfo {volume} {D54}},\
  \bibinfo {pages} {6233} (\bibinfo {year} {1996})},\ \Eprint
  {http://arxiv.org/abs/gr-qc/9602052} {arXiv:gr-qc/9602052 [gr-qc]}
  \BibitemShut {NoStop}%
\bibitem [{\citenamefont {Burgess}\ and\ \citenamefont
  {Williams}(2014)}]{Burgess:2014lwa}%
  \BibitemOpen
  \bibfield  {author} {\bibinfo {author} {\bibfnamefont {C.~P.}\ \bibnamefont
  {Burgess}}\ and\ \bibinfo {author} {\bibfnamefont {M.}~\bibnamefont
  {Williams}},\ }\href {\doibase 10.1007/JHEP08(2014)074} {\bibfield  {journal}
  {\bibinfo  {journal} {JHEP}\ }\textbf {\bibinfo {volume} {08}},\ \bibinfo
  {pages} {074} (\bibinfo {year} {2014})},\ \Eprint
  {http://arxiv.org/abs/1404.2236} {arXiv:1404.2236 [gr-qc]} \BibitemShut
  {NoStop}%
\bibitem [{\citenamefont {Solomon}\ and\ \citenamefont
  {Trodden}(2018)}]{Solomon:2017nlh}%
  \BibitemOpen
  \bibfield  {author} {\bibinfo {author} {\bibfnamefont {A.~R.}\ \bibnamefont
  {Solomon}}\ and\ \bibinfo {author} {\bibfnamefont {M.}~\bibnamefont
  {Trodden}},\ }\href {\doibase 10.1088/1475-7516/2018/02/031} {\bibfield
  {journal} {\bibinfo  {journal} {JCAP}\ }\textbf {\bibinfo {volume} {1802}},\
  \bibinfo {pages} {031} (\bibinfo {year} {2018})},\ \Eprint
  {http://arxiv.org/abs/1709.09695} {arXiv:1709.09695 [hep-th]} \BibitemShut
  {NoStop}%
\bibitem [{\citenamefont {Yunes}\ and\ \citenamefont
  {Pretorius}(2009)}]{Yunes:2009hc}%
  \BibitemOpen
  \bibfield  {author} {\bibinfo {author} {\bibfnamefont {N.}~\bibnamefont
  {Yunes}}\ and\ \bibinfo {author} {\bibfnamefont {F.}~\bibnamefont
  {Pretorius}},\ }\href {\doibase 10.1103/PhysRevD.79.084043} {\bibfield
  {journal} {\bibinfo  {journal} {Phys. Rev.}\ }\textbf {\bibinfo {volume}
  {D79}},\ \bibinfo {pages} {084043} (\bibinfo {year} {2009})},\ \Eprint
  {http://arxiv.org/abs/0902.4669} {arXiv:0902.4669 [gr-qc]} \BibitemShut
  {NoStop}%
\bibitem [{\citenamefont {Yagi}\ \emph {et~al.}(2012)\citenamefont {Yagi},
  \citenamefont {Stein}, \citenamefont {Yunes},\ and\ \citenamefont
  {Tanaka}}]{Yagi:2011xp}%
  \BibitemOpen
  \bibfield  {author} {\bibinfo {author} {\bibfnamefont {K.}~\bibnamefont
  {Yagi}}, \bibinfo {author} {\bibfnamefont {L.~C.}\ \bibnamefont {Stein}},
  \bibinfo {author} {\bibfnamefont {N.}~\bibnamefont {Yunes}}, \ and\ \bibinfo
  {author} {\bibfnamefont {T.}~\bibnamefont {Tanaka}},\ }\href {\doibase
  10.1103/PhysRevD.93.029902} {\bibfield  {journal} {\bibinfo  {journal} {Phys.
  Rev.}\ }\textbf {\bibinfo {volume} {D85}},\ \bibinfo {pages} {064022}
  (\bibinfo {year} {2012})},\ \bibinfo {note} {[Erratum: Phys.
  Rev.D93,no.2,029902(2016)]},\ \Eprint {http://arxiv.org/abs/1110.5950}
  {arXiv:1110.5950 [gr-qc]} \BibitemShut {NoStop}%
\bibitem [{\citenamefont {Stein}(2014)}]{Stein:2014xba}%
  \BibitemOpen
  \bibfield  {author} {\bibinfo {author} {\bibfnamefont {L.~C.}\ \bibnamefont
  {Stein}},\ }\href {\doibase 10.1103/PhysRevD.90.044061} {\bibfield  {journal}
  {\bibinfo  {journal} {Phys. Rev.}\ }\textbf {\bibinfo {volume} {D90}},\
  \bibinfo {pages} {044061} (\bibinfo {year} {2014})},\ \Eprint
  {http://arxiv.org/abs/1407.2350} {arXiv:1407.2350 [gr-qc]} \BibitemShut
  {NoStop}%
\bibitem [{\citenamefont {Bender}\ and\ \citenamefont
  {Orszag}(1978)}]{MR538168}%
  \BibitemOpen
  \bibfield  {author} {\bibinfo {author} {\bibfnamefont {C.~M.}\ \bibnamefont
  {Bender}}\ and\ \bibinfo {author} {\bibfnamefont {S.~A.}\ \bibnamefont
  {Orszag}},\ }\href@noop {} {\emph {\bibinfo {title} {Advanced mathematical
  methods for scientists and engineers}}}\ (\bibinfo  {publisher} {McGraw-Hill
  Book Co., New York},\ \bibinfo {year} {1978})\ pp.\ \bibinfo {pages}
  {xiv+593},\ \bibinfo {note} {international Series in Pure and Applied
  Mathematics}\BibitemShut {NoStop}%
\bibitem [{\citenamefont {Lovelace}(2009)}]{Lovelace:2008hd}%
  \BibitemOpen
  \bibfield  {author} {\bibinfo {author} {\bibfnamefont {G.}~\bibnamefont
  {Lovelace}},\ }\bibfield  {booktitle} {\emph {\bibinfo {booktitle}
  {{Numerical relativity data analysis. Proceedings, 2nd Meeting, NRDA 2008,
  Syracuse, USA, August 11-14, 2008}}},\ }\href {\doibase
  10.1088/0264-9381/26/11/114002} {\bibfield  {journal} {\bibinfo  {journal}
  {Class. Quant. Grav.}\ }\textbf {\bibinfo {volume} {26}},\ \bibinfo {pages}
  {114002} (\bibinfo {year} {2009})},\ \Eprint {http://arxiv.org/abs/0812.3132}
  {arXiv:0812.3132 [gr-qc]} \BibitemShut {NoStop}%
\bibitem [{\citenamefont {Pfeiffer}\ \emph {et~al.}(2007)\citenamefont
  {Pfeiffer}, \citenamefont {Brown}, \citenamefont {Kidder}, \citenamefont
  {Lindblom}, \citenamefont {Lovelace},\ and\ \citenamefont
  {Scheel}}]{Pfeiffer:2007yz}%
  \BibitemOpen
  \bibfield  {author} {\bibinfo {author} {\bibfnamefont {H.~P.}\ \bibnamefont
  {Pfeiffer}}, \bibinfo {author} {\bibfnamefont {D.~A.}\ \bibnamefont {Brown}},
  \bibinfo {author} {\bibfnamefont {L.~E.}\ \bibnamefont {Kidder}}, \bibinfo
  {author} {\bibfnamefont {L.}~\bibnamefont {Lindblom}}, \bibinfo {author}
  {\bibfnamefont {G.}~\bibnamefont {Lovelace}}, \ and\ \bibinfo {author}
  {\bibfnamefont {M.~A.}\ \bibnamefont {Scheel}},\ }\bibfield  {booktitle}
  {\emph {\bibinfo {booktitle} {{New frontiers in numerical relativity.
  Proceedings, International Meeting, NFNR 2006, Potsdam, Germany, July 17-21,
  2006}}},\ }\href {\doibase 10.1088/0264-9381/24/12/S06} {\bibfield  {journal}
  {\bibinfo  {journal} {Class. Quant. Grav.}\ }\textbf {\bibinfo {volume}
  {24}},\ \bibinfo {pages} {S59} (\bibinfo {year} {2007})},\ \Eprint
  {http://arxiv.org/abs/gr-qc/0702106} {arXiv:gr-qc/0702106 [gr-qc]}
  \BibitemShut {NoStop}%
\bibitem [{\citenamefont {Buonanno}\ \emph {et~al.}(2011)\citenamefont
  {Buonanno}, \citenamefont {Kidder}, \citenamefont {Mroue}, \citenamefont
  {Pfeiffer},\ and\ \citenamefont {Taracchini}}]{Buonanno:2010yk}%
  \BibitemOpen
  \bibfield  {author} {\bibinfo {author} {\bibfnamefont {A.}~\bibnamefont
  {Buonanno}}, \bibinfo {author} {\bibfnamefont {L.~E.}\ \bibnamefont
  {Kidder}}, \bibinfo {author} {\bibfnamefont {A.~H.}\ \bibnamefont {Mroue}},
  \bibinfo {author} {\bibfnamefont {H.~P.}\ \bibnamefont {Pfeiffer}}, \ and\
  \bibinfo {author} {\bibfnamefont {A.}~\bibnamefont {Taracchini}},\ }\href
  {\doibase 10.1103/PhysRevD.83.104034} {\bibfield  {journal} {\bibinfo
  {journal} {Phys. Rev.}\ }\textbf {\bibinfo {volume} {D83}},\ \bibinfo {pages}
  {104034} (\bibinfo {year} {2011})},\ \Eprint {http://arxiv.org/abs/1012.1549}
  {arXiv:1012.1549 [gr-qc]} \BibitemShut {NoStop}%
\bibitem [{\citenamefont {Hemberger}\ \emph {et~al.}(2013)\citenamefont
  {Hemberger}, \citenamefont {Scheel}, \citenamefont {Kidder}, \citenamefont
  {Szil{\'a}gyi}, \citenamefont {Lovelace}, \citenamefont {Taylor},\ and\
  \citenamefont {Teukolsky}}]{Hemberger:2012jz}%
  \BibitemOpen
  \bibfield  {author} {\bibinfo {author} {\bibfnamefont {D.~A.}\ \bibnamefont
  {Hemberger}}, \bibinfo {author} {\bibfnamefont {M.~A.}\ \bibnamefont
  {Scheel}}, \bibinfo {author} {\bibfnamefont {L.~E.}\ \bibnamefont {Kidder}},
  \bibinfo {author} {\bibfnamefont {B.}~\bibnamefont {Szil{\'a}gyi}}, \bibinfo
  {author} {\bibfnamefont {G.}~\bibnamefont {Lovelace}}, \bibinfo {author}
  {\bibfnamefont {N.~W.}\ \bibnamefont {Taylor}}, \ and\ \bibinfo {author}
  {\bibfnamefont {S.~A.}\ \bibnamefont {Teukolsky}},\ }\href {\doibase
  10.1088/0264-9381/30/11/115001} {\bibfield  {journal} {\bibinfo  {journal}
  {Class. Quant. Grav.}\ }\textbf {\bibinfo {volume} {30}},\ \bibinfo {pages}
  {115001} (\bibinfo {year} {2013})},\ \Eprint {http://arxiv.org/abs/1211.6079}
  {arXiv:1211.6079 [gr-qc]} \BibitemShut {NoStop}%
\bibitem [{\citenamefont {Abbott}\ \emph
  {et~al.}(2016{\natexlab{c}})\citenamefont {Abbott} \emph
  {et~al.}}]{TheLIGOScientific:2016wfe}%
  \BibitemOpen
  \bibfield  {author} {\bibinfo {author} {\bibfnamefont {B.~P.}\ \bibnamefont
  {Abbott}} \emph {et~al.} (\bibinfo {collaboration} {LIGO Scientific,
  Virgo}),\ }\href {\doibase 10.1103/PhysRevLett.116.241102} {\bibfield
  {journal} {\bibinfo  {journal} {Phys. Rev. Lett.}\ }\textbf {\bibinfo
  {volume} {116}},\ \bibinfo {pages} {241102} (\bibinfo {year}
  {2016}{\natexlab{c}})},\ \Eprint {http://arxiv.org/abs/1602.03840}
  {arXiv:1602.03840 [gr-qc]} \BibitemShut {NoStop}%
\bibitem [{\citenamefont {Kumar}\ \emph {et~al.}(2019)\citenamefont {Kumar},
  \citenamefont {Blackman}, \citenamefont {Field}, \citenamefont {Scheel},
  \citenamefont {Galley}, \citenamefont {Boyle}, \citenamefont {Kidder},
  \citenamefont {Pfeiffer}, \citenamefont {Szilagyi},\ and\ \citenamefont
  {Teukolsky}}]{Kumar:2018hml}%
  \BibitemOpen
  \bibfield  {author} {\bibinfo {author} {\bibfnamefont {P.}~\bibnamefont
  {Kumar}}, \bibinfo {author} {\bibfnamefont {J.}~\bibnamefont {Blackman}},
  \bibinfo {author} {\bibfnamefont {S.~E.}\ \bibnamefont {Field}}, \bibinfo
  {author} {\bibfnamefont {M.}~\bibnamefont {Scheel}}, \bibinfo {author}
  {\bibfnamefont {C.~R.}\ \bibnamefont {Galley}}, \bibinfo {author}
  {\bibfnamefont {M.}~\bibnamefont {Boyle}}, \bibinfo {author} {\bibfnamefont
  {L.~E.}\ \bibnamefont {Kidder}}, \bibinfo {author} {\bibfnamefont {H.~P.}\
  \bibnamefont {Pfeiffer}}, \bibinfo {author} {\bibfnamefont {B.}~\bibnamefont
  {Szilagyi}}, \ and\ \bibinfo {author} {\bibfnamefont {S.~A.}\ \bibnamefont
  {Teukolsky}},\ }\href {\doibase 10.1103/PhysRevD.99.124005} {\bibfield
  {journal} {\bibinfo  {journal} {Phys. Rev.}\ }\textbf {\bibinfo {volume}
  {D99}},\ \bibinfo {pages} {124005} (\bibinfo {year} {2019})},\ \Eprint
  {http://arxiv.org/abs/1808.08004} {arXiv:1808.08004 [gr-qc]} \BibitemShut
  {NoStop}%
\bibitem [{SXS()}]{SXSCatalog}%
  \BibitemOpen
  \href@noop {} {\enquote {\bibinfo {title} {{SXS Gravitational Waveform
  Database}},}\ }\bibinfo {howpublished}
  {\url{https://www.black-holes.org/waveforms}}\BibitemShut {NoStop}%
\bibitem [{\citenamefont {Lovelace}\ \emph {et~al.}(2016)\citenamefont
  {Lovelace} \emph {et~al.}}]{Lovelace:2016uwp}%
  \BibitemOpen
  \bibfield  {author} {\bibinfo {author} {\bibfnamefont {G.}~\bibnamefont
  {Lovelace}} \emph {et~al.},\ }\href {\doibase 10.1088/0264-9381/33/24/244002}
  {\bibfield  {journal} {\bibinfo  {journal} {Class. Quant. Grav.}\ }\textbf
  {\bibinfo {volume} {33}},\ \bibinfo {pages} {244002} (\bibinfo {year}
  {2016})},\ \Eprint {http://arxiv.org/abs/1607.05377} {arXiv:1607.05377
  [gr-qc]} \BibitemShut {NoStop}%
\bibitem [{\citenamefont {Bhagwat}\ \emph {et~al.}(2018)\citenamefont
  {Bhagwat}, \citenamefont {Okounkova}, \citenamefont {Ballmer}, \citenamefont
  {Brown}, \citenamefont {Giesler}, \citenamefont {Scheel},\ and\ \citenamefont
  {Teukolsky}}]{Bhagwat:2017tkm}%
  \BibitemOpen
  \bibfield  {author} {\bibinfo {author} {\bibfnamefont {S.}~\bibnamefont
  {Bhagwat}}, \bibinfo {author} {\bibfnamefont {M.}~\bibnamefont {Okounkova}},
  \bibinfo {author} {\bibfnamefont {S.~W.}\ \bibnamefont {Ballmer}}, \bibinfo
  {author} {\bibfnamefont {D.~A.}\ \bibnamefont {Brown}}, \bibinfo {author}
  {\bibfnamefont {M.}~\bibnamefont {Giesler}}, \bibinfo {author} {\bibfnamefont
  {M.~A.}\ \bibnamefont {Scheel}}, \ and\ \bibinfo {author} {\bibfnamefont
  {S.~A.}\ \bibnamefont {Teukolsky}},\ }\href {\doibase
  10.1103/PhysRevD.97.104065} {\bibfield  {journal} {\bibinfo  {journal} {Phys.
  Rev.}\ }\textbf {\bibinfo {volume} {D97}},\ \bibinfo {pages} {104065}
  (\bibinfo {year} {2018})},\ \Eprint {http://arxiv.org/abs/1711.00926}
  {arXiv:1711.00926 [gr-qc]} \BibitemShut {NoStop}%
\bibitem [{\citenamefont {Giesler}\ \emph {et~al.}(2019)\citenamefont
  {Giesler}, \citenamefont {Isi}, \citenamefont {Scheel},\ and\ \citenamefont
  {Teukolsky}}]{Giesler:2019uxc}%
  \BibitemOpen
  \bibfield  {author} {\bibinfo {author} {\bibfnamefont {M.}~\bibnamefont
  {Giesler}}, \bibinfo {author} {\bibfnamefont {M.}~\bibnamefont {Isi}},
  \bibinfo {author} {\bibfnamefont {M.}~\bibnamefont {Scheel}}, \ and\ \bibinfo
  {author} {\bibfnamefont {S.}~\bibnamefont {Teukolsky}},\ }\href@noop {} {\
  (\bibinfo {year} {2019})},\ \Eprint {http://arxiv.org/abs/1903.08284}
  {arXiv:1903.08284 [gr-qc]} \BibitemShut {NoStop}%
\bibitem [{\citenamefont {Varma}\ \emph {et~al.}(2019)\citenamefont {Varma},
  \citenamefont {Field}, \citenamefont {Scheel}, \citenamefont {Blackman},
  \citenamefont {Kidder},\ and\ \citenamefont {Pfeiffer}}]{Varma:2018mmi}%
  \BibitemOpen
  \bibfield  {author} {\bibinfo {author} {\bibfnamefont {V.}~\bibnamefont
  {Varma}}, \bibinfo {author} {\bibfnamefont {S.~E.}\ \bibnamefont {Field}},
  \bibinfo {author} {\bibfnamefont {M.~A.}\ \bibnamefont {Scheel}}, \bibinfo
  {author} {\bibfnamefont {J.}~\bibnamefont {Blackman}}, \bibinfo {author}
  {\bibfnamefont {L.~E.}\ \bibnamefont {Kidder}}, \ and\ \bibinfo {author}
  {\bibfnamefont {H.~P.}\ \bibnamefont {Pfeiffer}},\ }\href {\doibase
  10.1103/PhysRevD.99.064045} {\bibfield  {journal} {\bibinfo  {journal} {Phys.
  Rev.}\ }\textbf {\bibinfo {volume} {D99}},\ \bibinfo {pages} {064045}
  (\bibinfo {year} {2019})},\ \Eprint {http://arxiv.org/abs/1812.07865}
  {arXiv:1812.07865 [gr-qc]} \BibitemShut {NoStop}%
\bibitem [{\citenamefont {Shoemaker}(2011)}]{LIGOPSD}%
  \BibitemOpen
  \bibfield  {author} {\bibinfo {author} {\bibfnamefont {D.}~\bibnamefont
  {Shoemaker}} (\bibinfo {collaboration} {LIGO Scientific, VIRGO}),\
  }\href@noop {} {\enquote {\bibinfo {title} {Advanced ligo anticipated
  sensitivity curves},}\ }\bibinfo {howpublished}
  {\url{https://dcc.ligo.org/LIGO-T0900288/public}} (\bibinfo {year}
  {2011})\BibitemShut {NoStop}%
\bibitem [{\citenamefont {Chatziioannou}\ \emph {et~al.}(2017)\citenamefont
  {Chatziioannou}, \citenamefont {Klein}, \citenamefont {Yunes},\ and\
  \citenamefont {Cornish}}]{Chatziioannou:2017tdw}%
  \BibitemOpen
  \bibfield  {author} {\bibinfo {author} {\bibfnamefont {K.}~\bibnamefont
  {Chatziioannou}}, \bibinfo {author} {\bibfnamefont {A.}~\bibnamefont
  {Klein}}, \bibinfo {author} {\bibfnamefont {N.}~\bibnamefont {Yunes}}, \ and\
  \bibinfo {author} {\bibfnamefont {N.}~\bibnamefont {Cornish}},\ }\href
  {\doibase 10.1103/PhysRevD.95.104004} {\bibfield  {journal} {\bibinfo
  {journal} {Phys. Rev.}\ }\textbf {\bibinfo {volume} {D95}},\ \bibinfo {pages}
  {104004} (\bibinfo {year} {2017})},\ \Eprint
  {http://arxiv.org/abs/1703.03967} {arXiv:1703.03967 [gr-qc]} \BibitemShut
  {NoStop}%
\bibitem [{\citenamefont {Taracchini}\ \emph {et~al.}(2014)\citenamefont
  {Taracchini} \emph {et~al.}}]{Taracchini:2013rva}%
  \BibitemOpen
  \bibfield  {author} {\bibinfo {author} {\bibfnamefont {A.}~\bibnamefont
  {Taracchini}} \emph {et~al.},\ }\href {\doibase 10.1103/PhysRevD.89.061502}
  {\bibfield  {journal} {\bibinfo  {journal} {Phys. Rev.}\ }\textbf {\bibinfo
  {volume} {D89}},\ \bibinfo {pages} {061502} (\bibinfo {year} {2014})},\
  \Eprint {http://arxiv.org/abs/1311.2544} {arXiv:1311.2544 [gr-qc]}
  \BibitemShut {NoStop}%
\bibitem [{\citenamefont {Khan}\ \emph {et~al.}(2016)\citenamefont {Khan},
  \citenamefont {Husa}, \citenamefont {Hannam}, \citenamefont {Ohme},
  \citenamefont {Pürrer}, \citenamefont {Jiménez~Forteza},\ and\
  \citenamefont {Bohé}}]{Khan:2015jqa}%
  \BibitemOpen
  \bibfield  {author} {\bibinfo {author} {\bibfnamefont {S.}~\bibnamefont
  {Khan}}, \bibinfo {author} {\bibfnamefont {S.}~\bibnamefont {Husa}}, \bibinfo
  {author} {\bibfnamefont {M.}~\bibnamefont {Hannam}}, \bibinfo {author}
  {\bibfnamefont {F.}~\bibnamefont {Ohme}}, \bibinfo {author} {\bibfnamefont
  {M.}~\bibnamefont {Pürrer}}, \bibinfo {author} {\bibfnamefont
  {X.}~\bibnamefont {Jiménez~Forteza}}, \ and\ \bibinfo {author}
  {\bibfnamefont {A.}~\bibnamefont {Bohé}},\ }\href {\doibase
  10.1103/PhysRevD.93.044007} {\bibfield  {journal} {\bibinfo  {journal} {Phys.
  Rev.}\ }\textbf {\bibinfo {volume} {D93}},\ \bibinfo {pages} {044007}
  (\bibinfo {year} {2016})},\ \Eprint {http://arxiv.org/abs/1508.07253}
  {arXiv:1508.07253 [gr-qc]} \BibitemShut {NoStop}%
\bibitem [{\citenamefont {Pürrer}\ and\ \citenamefont
  {Haster}(2019)}]{Purrer:2019jcp}%
  \BibitemOpen
  \bibfield  {author} {\bibinfo {author} {\bibfnamefont {M.}~\bibnamefont
  {Pürrer}}\ and\ \bibinfo {author} {\bibfnamefont {C.-J.}\ \bibnamefont
  {Haster}},\ }\href@noop {} {\  (\bibinfo {year} {2019})},\ \Eprint
  {http://arxiv.org/abs/1912.10055} {arXiv:1912.10055 [gr-qc]} \BibitemShut
  {NoStop}%
\bibitem [{\citenamefont {Isi}\ \emph {et~al.}(2019)\citenamefont {Isi},
  \citenamefont {Giesler}, \citenamefont {Farr}, \citenamefont {Scheel},\ and\
  \citenamefont {Teukolsky}}]{Isi:2019aib}%
  \BibitemOpen
  \bibfield  {author} {\bibinfo {author} {\bibfnamefont {M.}~\bibnamefont
  {Isi}}, \bibinfo {author} {\bibfnamefont {M.}~\bibnamefont {Giesler}},
  \bibinfo {author} {\bibfnamefont {W.~M.}\ \bibnamefont {Farr}}, \bibinfo
  {author} {\bibfnamefont {M.~A.}\ \bibnamefont {Scheel}}, \ and\ \bibinfo
  {author} {\bibfnamefont {S.~A.}\ \bibnamefont {Teukolsky}},\ }\href {\doibase
  10.1103/PhysRevLett.123.111102} {\bibfield  {journal} {\bibinfo  {journal}
  {Phys. Rev. Lett.}\ }\textbf {\bibinfo {volume} {123}},\ \bibinfo {pages}
  {111102} (\bibinfo {year} {2019})},\ \Eprint
  {http://arxiv.org/abs/1905.00869} {arXiv:1905.00869 [gr-qc]} \BibitemShut
  {NoStop}%
\bibitem [{\citenamefont {Aasi}\ \emph {et~al.}(2013)\citenamefont {Aasi} \emph
  {et~al.}}]{Aasi:2013jjl}%
  \BibitemOpen
  \bibfield  {author} {\bibinfo {author} {\bibfnamefont {J.}~\bibnamefont
  {Aasi}} \emph {et~al.} (\bibinfo {collaboration} {LIGO Scientific, VIRGO}),\
  }\href {\doibase 10.1103/PhysRevD.88.062001} {\bibfield  {journal} {\bibinfo
  {journal} {Phys. Rev.}\ }\textbf {\bibinfo {volume} {D88}},\ \bibinfo {pages}
  {062001} (\bibinfo {year} {2013})},\ \Eprint {http://arxiv.org/abs/1304.1775}
  {arXiv:1304.1775 [gr-qc]} \BibitemShut {NoStop}%
\bibitem [{\citenamefont {Cornish}\ and\ \citenamefont
  {Littenberg}(2015)}]{Cornish:2014kda}%
  \BibitemOpen
  \bibfield  {author} {\bibinfo {author} {\bibfnamefont {N.~J.}\ \bibnamefont
  {Cornish}}\ and\ \bibinfo {author} {\bibfnamefont {T.~B.}\ \bibnamefont
  {Littenberg}},\ }\href {\doibase 10.1088/0264-9381/32/13/135012} {\bibfield
  {journal} {\bibinfo  {journal} {Class. Quant. Grav.}\ }\textbf {\bibinfo
  {volume} {32}},\ \bibinfo {pages} {135012} (\bibinfo {year} {2015})},\
  \Eprint {http://arxiv.org/abs/1410.3835} {arXiv:1410.3835 [gr-qc]}
  \BibitemShut {NoStop}%
\bibitem [{\citenamefont {Witek}\ \emph {et~al.}(2019)\citenamefont {Witek},
  \citenamefont {Gualtieri}, \citenamefont {Pani},\ and\ \citenamefont
  {Sotiriou}}]{Witek:2018dmd}%
  \BibitemOpen
  \bibfield  {author} {\bibinfo {author} {\bibfnamefont {H.}~\bibnamefont
  {Witek}}, \bibinfo {author} {\bibfnamefont {L.}~\bibnamefont {Gualtieri}},
  \bibinfo {author} {\bibfnamefont {P.}~\bibnamefont {Pani}}, \ and\ \bibinfo
  {author} {\bibfnamefont {T.~P.}\ \bibnamefont {Sotiriou}},\ }\href {\doibase
  10.1103/PhysRevD.99.064035} {\bibfield  {journal} {\bibinfo  {journal} {Phys.
  Rev.}\ }\textbf {\bibinfo {volume} {D99}},\ \bibinfo {pages} {064035}
  (\bibinfo {year} {2019})},\ \Eprint {http://arxiv.org/abs/1810.05177}
  {arXiv:1810.05177 [gr-qc]} \BibitemShut {NoStop}%
\bibitem [{\citenamefont {Okounkova}(2019)}]{Okounkova:2019zep}%
  \BibitemOpen
  \bibfield  {author} {\bibinfo {author} {\bibfnamefont {M.}~\bibnamefont
  {Okounkova}},\ }\href@noop {} {\  (\bibinfo {year} {2019})},\ \Eprint
  {http://arxiv.org/abs/1909.12251} {arXiv:1909.12251 [gr-qc]} \BibitemShut
  {NoStop}%
\bibitem [{SpE()}]{SpECwebsite}%
  \BibitemOpen
  \href@noop {} {\enquote {\bibinfo {title} {The {S}pectral {E}instein {C}ode
  ({SpEC})},}\ }\bibinfo {howpublished}
  {\url{http://www.black-holes.org/SpEC.html}}\BibitemShut {NoStop}%
\bibitem [{\citenamefont {Galley}\ and\ \citenamefont
  {Rothstein}(2017)}]{Galley:2016zee}%
  \BibitemOpen
  \bibfield  {author} {\bibinfo {author} {\bibfnamefont {C.~R.}\ \bibnamefont
  {Galley}}\ and\ \bibinfo {author} {\bibfnamefont {I.~Z.}\ \bibnamefont
  {Rothstein}},\ }\href {\doibase 10.1103/PhysRevD.95.104054} {\bibfield
  {journal} {\bibinfo  {journal} {Phys. Rev.}\ }\textbf {\bibinfo {volume}
  {D95}},\ \bibinfo {pages} {104054} (\bibinfo {year} {2017})},\ \Eprint
  {http://arxiv.org/abs/1609.08268} {arXiv:1609.08268 [gr-qc]} \BibitemShut
  {NoStop}%
\bibitem [{\citenamefont {Kevorkian}\ and\ \citenamefont
  {Cole}(1996)}]{Kevorkian}%
  \BibitemOpen
  \bibfield  {author} {\bibinfo {author} {\bibfnamefont {J.}~\bibnamefont
  {Kevorkian}}\ and\ \bibinfo {author} {\bibfnamefont {J.~D.}\ \bibnamefont
  {Cole}},\ }\href@noop {} {\emph {\bibinfo {title} {Multiple Scale and
  Singular Perturbation Methods}}}\ (\bibinfo  {publisher} {Springer},\
  \bibinfo {address} {New York},\ \bibinfo {year} {1996})\BibitemShut {NoStop}%
\bibitem [{\citenamefont {Kunihiro}(1995)}]{Kunihiro:1995zt}%
  \BibitemOpen
  \bibfield  {author} {\bibinfo {author} {\bibfnamefont {T.}~\bibnamefont
  {Kunihiro}},\ }\href {\doibase 10.1143/PTP.94.503} {\bibfield  {journal}
  {\bibinfo  {journal} {Prog. Theor. Phys.}\ }\textbf {\bibinfo {volume}
  {94}},\ \bibinfo {pages} {503} (\bibinfo {year} {1995})},\ \bibinfo {note}
  {[Erratum: Prog. Theor. Phys.95,835(1996)]},\ \Eprint
  {http://arxiv.org/abs/hep-th/9505166} {arXiv:hep-th/9505166 [hep-th]}
  \BibitemShut {NoStop}%
\bibitem [{\citenamefont {Will}\ and\ \citenamefont
  {Maitra}(2017)}]{Will:2016pgm}%
  \BibitemOpen
  \bibfield  {author} {\bibinfo {author} {\bibfnamefont {C.~M.}\ \bibnamefont
  {Will}}\ and\ \bibinfo {author} {\bibfnamefont {M.}~\bibnamefont {Maitra}},\
  }\href {\doibase 10.1103/PhysRevD.95.064003} {\bibfield  {journal} {\bibinfo
  {journal} {Phys. Rev.}\ }\textbf {\bibinfo {volume} {D95}},\ \bibinfo {pages}
  {064003} (\bibinfo {year} {2017})},\ \Eprint
  {http://arxiv.org/abs/1611.06931} {arXiv:1611.06931 [gr-qc]} \BibitemShut
  {NoStop}%
\bibitem [{\citenamefont {Pound}\ \emph {et~al.}(2005)\citenamefont {Pound},
  \citenamefont {Poisson},\ and\ \citenamefont {Nickel}}]{Pound:2005fs}%
  \BibitemOpen
  \bibfield  {author} {\bibinfo {author} {\bibfnamefont {A.}~\bibnamefont
  {Pound}}, \bibinfo {author} {\bibfnamefont {E.}~\bibnamefont {Poisson}}, \
  and\ \bibinfo {author} {\bibfnamefont {B.~G.}\ \bibnamefont {Nickel}},\
  }\href {\doibase 10.1103/PhysRevD.72.124001} {\bibfield  {journal} {\bibinfo
  {journal} {Phys. Rev.}\ }\textbf {\bibinfo {volume} {D72}},\ \bibinfo {pages}
  {124001} (\bibinfo {year} {2005})},\ \Eprint
  {http://arxiv.org/abs/gr-qc/0509122} {arXiv:gr-qc/0509122 [gr-qc]}
  \BibitemShut {NoStop}%
\bibitem [{\citenamefont {Pound}\ and\ \citenamefont
  {Poisson}(2008)}]{Pound:2007th}%
  \BibitemOpen
  \bibfield  {author} {\bibinfo {author} {\bibfnamefont {A.}~\bibnamefont
  {Pound}}\ and\ \bibinfo {author} {\bibfnamefont {E.}~\bibnamefont
  {Poisson}},\ }\href {\doibase 10.1103/PhysRevD.77.044013} {\bibfield
  {journal} {\bibinfo  {journal} {Phys. Rev.}\ }\textbf {\bibinfo {volume}
  {D77}},\ \bibinfo {pages} {044013} (\bibinfo {year} {2008})},\ \Eprint
  {http://arxiv.org/abs/0708.3033} {arXiv:0708.3033 [gr-qc]} \BibitemShut
  {NoStop}%
\bibitem [{\citenamefont {Warburton}\ \emph {et~al.}(2012)\citenamefont
  {Warburton}, \citenamefont {Akcay}, \citenamefont {Barack}, \citenamefont
  {Gair},\ and\ \citenamefont {Sago}}]{Warburton:2011fk}%
  \BibitemOpen
  \bibfield  {author} {\bibinfo {author} {\bibfnamefont {N.}~\bibnamefont
  {Warburton}}, \bibinfo {author} {\bibfnamefont {S.}~\bibnamefont {Akcay}},
  \bibinfo {author} {\bibfnamefont {L.}~\bibnamefont {Barack}}, \bibinfo
  {author} {\bibfnamefont {J.~R.}\ \bibnamefont {Gair}}, \ and\ \bibinfo
  {author} {\bibfnamefont {N.}~\bibnamefont {Sago}},\ }\href {\doibase
  10.1103/PhysRevD.85.061501} {\bibfield  {journal} {\bibinfo  {journal} {Phys.
  Rev.}\ }\textbf {\bibinfo {volume} {D85}},\ \bibinfo {pages} {061501}
  (\bibinfo {year} {2012})},\ \Eprint {http://arxiv.org/abs/1111.6908}
  {arXiv:1111.6908 [gr-qc]} \BibitemShut {NoStop}%
\bibitem [{\citenamefont {Hinderer}\ and\ \citenamefont
  {Flanagan}(2008)}]{Hinderer:2008dm}%
  \BibitemOpen
  \bibfield  {author} {\bibinfo {author} {\bibfnamefont {T.}~\bibnamefont
  {Hinderer}}\ and\ \bibinfo {author} {\bibfnamefont {E.~E.}\ \bibnamefont
  {Flanagan}},\ }\href {\doibase 10.1103/PhysRevD.78.064028} {\bibfield
  {journal} {\bibinfo  {journal} {Phys. Rev.}\ }\textbf {\bibinfo {volume}
  {D78}},\ \bibinfo {pages} {064028} (\bibinfo {year} {2008})},\ \Eprint
  {http://arxiv.org/abs/0805.3337} {arXiv:0805.3337 [gr-qc]} \BibitemShut
  {NoStop}%
\bibitem [{\citenamefont {Pound}(2015)}]{Pound:2015wva}%
  \BibitemOpen
  \bibfield  {author} {\bibinfo {author} {\bibfnamefont {A.}~\bibnamefont
  {Pound}},\ }\href {\doibase 10.1103/PhysRevD.92.104047} {\bibfield  {journal}
  {\bibinfo  {journal} {Phys. Rev.}\ }\textbf {\bibinfo {volume} {D92}},\
  \bibinfo {pages} {104047} (\bibinfo {year} {2015})},\ \Eprint
  {http://arxiv.org/abs/1510.05172} {arXiv:1510.05172 [gr-qc]} \BibitemShut
  {NoStop}%
\end{thebibliography}%
\end{document}